\documentclass[caps]{cupbook}

\begin{document}

\setcounter{chapter}{2}

\chapter[3D Instrumentation / WS XVII / Bershady]
{3D Spectroscopic Instrumentation  \\ \vspace{0.5cm}
\Large Matthew A. Bershady \\ Department of Astronomy,
University of Wisconsin}

In this Chapter\footnote{to appear in ``3D Spectroscopy in Astronomy,
  XVII Canary Island Winter School of Astrophysics,''
  eds. E. Mediavilla, S. Arribas, M. Roth, J. Cepa-Nogue, and
  F. Sanchez, Cambridge University Press, 2009.} we review the
challenges of, and opportunities for, 3D spectroscopy, and how these
have lead to new and different approaches to sampling astronomical
information. We describe and categorize existing instruments on 4m and
10m telescopes.  Our primary focus is on grating-dispersed
spectrographs. We discuss how to optimize dispersive elements, such as
VPH gratings, to achieve adequate spectral resolution, high
throughput, and efficient data packing to maximize spatial sampling
for 3D spectroscopy. We review and compare the various coupling
methods that make these spectrographs ``3D,'' including fibers,
lenslets, slicers, and filtered multi-slits.  We also describe
Fabry-Perot and spatial-heterodyne interferometers, pointing out their
advantages as field-widened systems relative to conventional,
grating-dispersed spectrographs. We explore the parameter space all
these instruments sample, highlighting regimes open for exploitation.
Present instruments provide a foil for future development. We give an
overview of plans for such future instruments on today's large
telescopes, in space, and in the coming era of extremely large
telescopes. Currently-planned instruments open new domains, but also
leave significant areas of parameter space vacant, beckoning further
development.

\section{Fundamental Challenges and Considerations}

\subsection{The Detector Limit-I: Six into Two Dimensions}

Astronomical data exist within 6-dimensional hyper-cube sampling two
spatial dimensions, one spectral dimension, one temporal dimension,
and two polarizations. In contrast, high-efficiency, panoramic digital
detectors today are only two-dimensional (with some limited
exceptions). The instrument-builder's trick is to down-select the
critical observational dimensions relevant to address a well-motivated
subset of science problems. Here we consider the application to 3D
spectroscopy at high photon count-rates, where both spatial and
spectral domains must be parsed onto, e.g., a CCD detector, as
illustrated in Figure 1.1. The choice is in how the data-cube is
sliced along orthogonal dimensions, since it isn't easy to rotate a
slice within the cube. Such ``rotation'' could be accomplished via
multi-fiber or multi-slicer feeds to multiple spectrographs, but to
date the science motivation has not led to such a design. In practice,
then, we have the extremes of single-object, cross-dispersed echelle
spectrographs, to Fabry-Perot (F-P) monochromators. The
``traditional'' integral-field spectrograph (IFS) is between these two
limiting domains.

\begin{figure}
\centering
\vspace{9cm}
\includegraphics{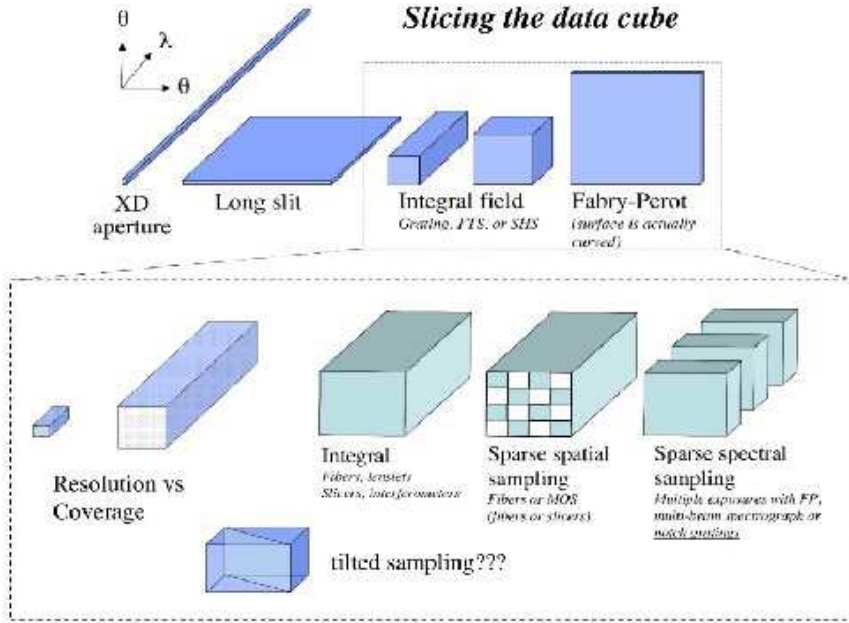}
\caption{Sampling the data-cube with equal volumes and detector
elements.}
\end{figure}

In addition to balancing the trade-offs between spatial versus spectral
information, there is also the issue of balancing sampling (i.e.,
resolution) versus coverage in either of these
dimensions. Science-driven trades formulate any specific instrument
design. When sampling spatial and spectral domains, not all data has
equal information content. Hence one may also consider integral versus
sparse sampling. Fiber-fed IFS such as Hexaflex (Arribas, Mediavilla
\& Rasilla 1991) and SparsePak (Bershady et al. 2004) are examples of
sparse-sampling in the spatial domain. Multi-exposure Fabry-Perot
observations, multi-beam spectrographs, or notch-gratings (discussed
below) are examples of instruments with the capability of sparse
sampling in the spectral domain.

\subsection{Merit Functions}

There are a number of generic merit functions found in the
instrumentation literature, in a variety of guises used, or tailored,
to suit the need of comparing or contrasting the niche of specific
instruments.  Some useful preliminary definitions (used throughout
this Chapter) are the spectral resolution, $R = \lambda/d\lambda$; the
number of spectral resolution elements, $N_R$; spectral coverage =
$\Delta\lambda = N_R \times d\lambda$; spatial resolution $d\Omega$,
i.e., the sampling element on the sky (fiber, lenslet, slicer
slit-let, or seeing-disk); number of spatial resolution elements,
N$_\Omega$; and spatial coverage $\Omega = N_\Omega \times d\Omega$.

With these definitions, the trade-offs discussed above may be
summarized by stating that $N_R \times N_\Omega$ must be roughly
constant for a given detector. Another important statement is that
$A \times \Omega$, or grasp, is conserved in an optical system ($A$ is the
telescope collecting area): The same instrument has the same $A\times\Omega$
on any diameter telescope with the same focal ratio -- something
derived from the identify $\Omega = a/f^2$, where $a$ is the
instrument focal area and $f$ the focal-length. What changes with
aperture, of course, is the angular sampling. For sufficiently
extended sources, angular sampling is not necessarily at a premium.
Imagine, for example, dissecting nearby galaxies with a MUSE-like
instrument on a 4m or 1m-class telescope. (MUSE is discussed later in
this Chapter; Bacon et al. 2004).

In addition to the basic ingredients listed above, the most common
merit functions are the grasp, the specific grasp, $ A \times d\Omega$
(how much is grasped within each spatial resolution element of the
instrument), and etendue, $A \times \Omega \times \epsilon$, where
$\epsilon$ is the total system efficiency from the top of the
atmosphere to the detected photo-electron. Etendue is more fundamental
than grasp since high-efficiency instruments are the true performance
engines.  Despite the fact that an instrument with an un-reported
efficiency is much like a car {\it sans} fuel-gauge or speedometer,
recovering $\epsilon$ from the literature is often not possible. For
this reason we resort to grasp, but note that in some
cases this gives an unfair comparison between instruments.

If there is no premium on spatial information then ``spectral power,''
$R \times N_R$, is suitable. At the opposite extreme, where spatial
information is paramount, a suitable merit function is $A \times
d\Omega^n \times N_\Omega = A \times d\Omega^{n-1} \times \Omega$,
where n = 1 for high specific grasp and -1 for high resolution.  In
the context of 3D spectroscopy, merit functions which combine spatial
and spectral power are appropriate: $\Omega \times R$, $A \times
\Omega \times R \times \epsilon$, or their counterparts replacing
$\Omega$ with $d\Omega$. If any information will do, $N_R \times N_\Omega$
alone gives a good synopsis of the instrument power since this
effectively gives the number of resolution elements (related to
detector elements) that have been effectively utilized by the
instrument.

An attempt at a grand merit function can be formulated by asking the
following, sweeping question: {\it How many resolution elements can be
coupled efficiently to the largest telescope aperture (A) covering the
largest patrol field ($\Omega_s$) for as little cost as possible?} In
this case, the figure of merit may be written: $$F.O.M. = \epsilon
\times (\Delta\lambda / \lambda) \times (\Omega/d\Omega) \times A
\times \Omega_s \times \pounds^{-1} \ = \ \epsilon \times N_R \times
N_\Omega \times A \times \Omega_s \times \pounds^{-1}$$ where
$\Delta\lambda$ is the sampled spectral range, and \pounds\ is the
cost in the suitable local currency. To this figure of merit one may
add the product $R^n \times d\Omega^m$, where $n,m = 1$ if resolution is
science-critical in the spectral and spatial domains (respectively),
$n,m = -1$ if coverage is science-critical, or $n,m = 0$ if resolution
and coverage are science-neutral (in which case you're not trying hard
enough!).

From this discussion it is clear that a suitable choice of merit
function is complicated, and {\it must} be science driven.  The
relative evaluation of instruments cannot be done sensibly in the
absence of a science-formulated F.O.M.; the outcome of any sensible
evaluation will therefore depend on the science-formulation. For this
reason, when we compare instruments we strategically retreat and
explore the multi-dimensional space of the fundamental parameters of
spatial resolution, spectral resolution, specific grasp, total grasp,
spectral power, and N$_R$ versus N$_\Omega$.

\subsection{Why Spectral Resolution is so Important}

\begin{figure}
\centering
\vspace{7cm}
\includegraphics{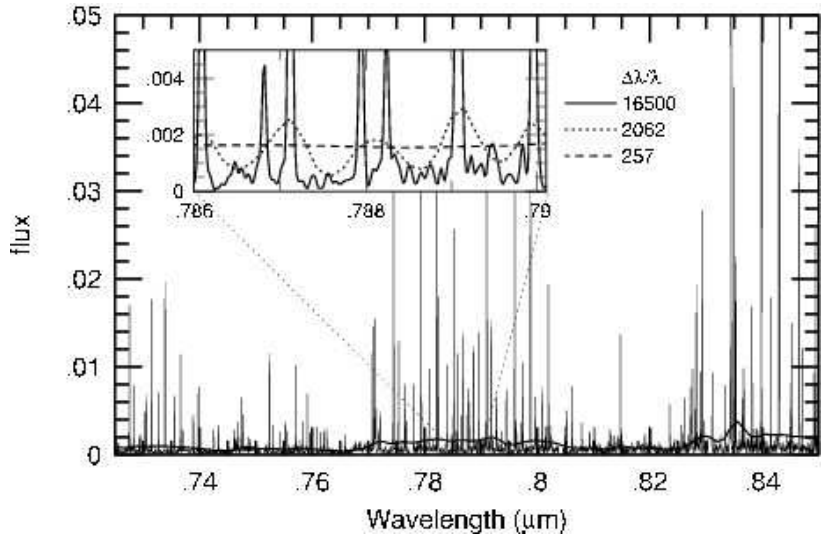}
\caption{Night Sky near 0.8 $\mu$m at $250 < R < 33,000$.}
\end{figure}

In addition to the intrinsic merits and requirement of high spectral
resolution for certain science programs, high resolution is of general
importance for improving signal-to-noise ($S/N$) in the red and
near-infrared.  For ground-based observations, terrestrial backgrounds
from 0.7-2.2 microns suffer a common malady of being dominated by
extremely narrow (m s$^{-1}$) air-glow lines, typically from OH
molecules. Unlike the thermal IR, however, there is a cure to lower
the background without going to high-altitude or space. The air-glow
lines cluster in bands, and the lines within the bands may be
separated at $R = $3000-5000. This means that at these resolutions,
while the mean background level within the spectral band-pass is
constant, the median drops precipitously: more spectral resolution
elements are at lower background level in inter-line regions. The
lines themselves, however, remain unresolved until $R \sim$ few
$\times \ 10^5$, so that above $R = 4000$ one continues to increase
the fraction of the spectral band-pass at low-background levels.

\begin{figure}
\centering
\vspace{7cm}
\includegraphics{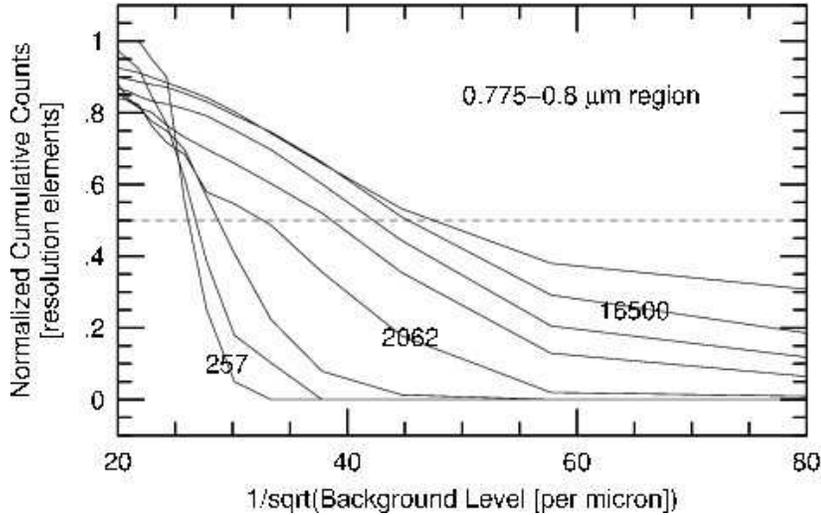}
\caption{Cumulative distribution of resolution elements as a function
of the background level proportional to $S/N$ (increasing to left) for
$250 < R < 33,000$ (labeled).}
\end{figure}

As an illustration, we show the terrestrial sky bacgkround in a
spectral region at 0.8 microns observed by D. York and J. Lauroesch
(private communication) with the KPNO 4m echelle. In Figure 1.2 the
sky spectra, observed at an instrumental resolution of 33,000, is
degraded to illustrate the resulting change in the distribution of
background levels.  In Figure 1.3, the normalized, cumulative
distribution of resolution elements as a function of background level
are plotted for different instrumental resolutions. For
background-limited measurements, the $S/N$ is proportional to the
inverse square-root of the background level.  Hence the median
background level gives an effective scaling for sensitivity gains with
spectral resolution. It can be seen the largest changes in the median
background level occur between $1000<R<4000$, but significant gains
continue at higher resolution.  The result can be qualitatively
generalized to other wavelengths in the 0.7-2.2 $\mu$m regime.  While
the lines become more intense moving to longer wavelengths, the
power-spectrum (in wavelength) of the lines appears roughly
independent of wavelength in this regime ({\it cf.} Maihara et
al. 1993 and Hanuschik 2003). Note this is a qualitative assessment
that should be formally quantified.

\subsection{The Detector Limit-II: Read-noise}

Our infatuation with spectral resolution is a problem given the modern
predilection for high angular resolution. After the Hubble Space Telescope
there is no turning back! There is, however, a limit, due to detector
noise, which we always want to be above. The goal is to be
photon-limited (either source or background) because this is
fundamental (it's the best we can do), and for practical purposes, $S/N$
is independent of sub-exposure time and detector sampling.

\begin{figure}
\centering
\vspace{8cm}
\includegraphics{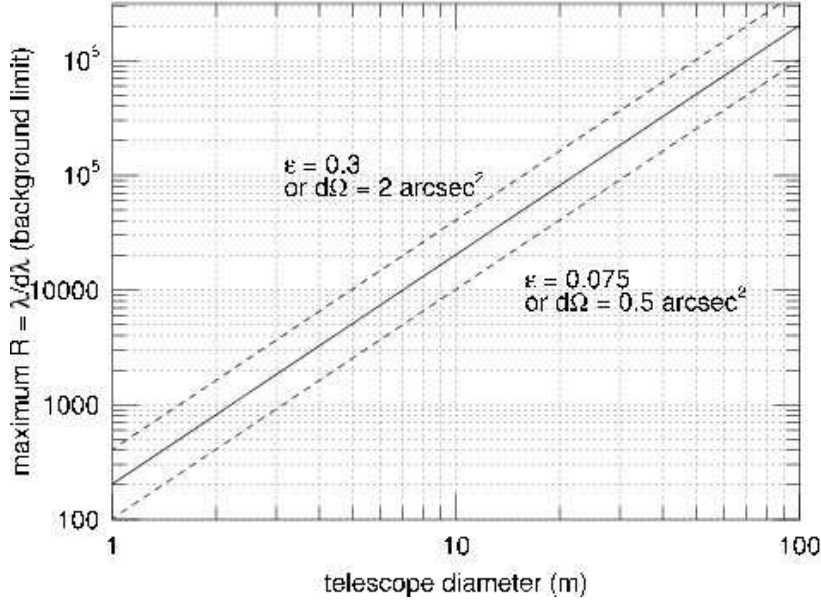}
\caption{Maximum spectral resolution versus telescope diameter to stay
background (vs detector) limited for different assumptions of
instrument efficiency ($\epsilon$) and spatial sampling
($d\Omega$). The solid line assumes $\epsilon = 0.15$ and $d\Omega = 1$
arcsec$^2$.}
\vspace{-0.25cm}
\end{figure}

To stay photon-limited in the background-limited regime puts
significant constraints on the $\Omega$-R sampling unit.  The spatial
and spectral sampling unit can't be too fine for a given $A$ and
$\epsilon$. For 8m- and 4m-class telescopes we calculate
\begin{eqnarray*}
R/d\Omega & < & 16500 (D_T/9m)^2 (t/1h) (\epsilon/0.15) \ arcsec^{-2}, \ {\rm or} \\
          & < & 2500 (D_T/3.5m)^2 (t/1h) (\epsilon/0.15) \ arcsec^{-2}, 
\end{eqnarray*}
where $D_T$ is the telescope aperture diameter and $t$ the
(single destructive-read) exposure length. The general case is shown
in Figure 1.4. To reach spectral resolutions well above $R = 5000$,
which is advantageous for background-reduction, a telescope
significantly in excess of 10m is needed for apertures significantly
under 1 arcsec$^{-2}$.

\smallskip
With these considerations in mind, in the next three sections (\S
1.2-1.4) we turn to approaches and examples of existing instruments,
followed by three sections (\S 1.5-1.7) in which we summarize the
range of these instruments, what parameter space is under-sampled, and
the prospects for future instruments. Throughout, we attempt to
provide relatively complete instrument lists. No doubt some
instruments have been over-looked, plus the field of instrumentation
advances rapidly. Reports of additional instruments or corrections
are welcome.\footnote{Send email to: mab@astro.wisc.edu.}

\section{Grating-Dispersed Spectrographs}

Basic spectrograph theory and design can be found in most standard
optics textbooks. Of particular note is the excellent monograph on
astronomical optics by Schroeder (2000). In \S 1.2.1 we summarize the
salient features to provide a consistent nomenclature, and to put
these features into context of our discussion of 3D spectroscopy,
specifically what drives consideration of merit functions that tune
spatial versus spectral performance. The balance of this section
includes a description of dispersive elements (\S 1.2.2), coupling
methods and modes (\S 1.2.3-11), and summary considerations --
including a discussion of sky-subtraction problems and solutions (\S
1.2.12).

\subsection{Basic Spectrograph Design}

In a 3D spectrographic system, there is a premium on packing spatial
information onto the detector. To achieve sufficient spectral
resolution at the same time requires balancing the trades between
system magnification and dispersion. Starting with the grating
equation, generalized for a grating immersed in medium of index $n$:
$m \ \lambda = n \ \Lambda_g ( sin \ \beta + sin \ \alpha )$, where
$\Lambda_g$ is the projected groove separation in the plane of the
grating, $m$ the order, and $\alpha$ and $\beta$ the incident and
diffracted grating angles relative to the grating normal in the
medium, we can write the angular and linear dispersion as $\gamma
\equiv d\beta/d\lambda = m \ / \ n \ \Lambda_g \ cos \ \beta = (sin \
\beta + sin \ \alpha) \ / \ \lambda \ cos \ \beta$, and $dl/d\lambda =
f_2 \gamma$. Figure 1.5 illustrates a basic spectrograph, defines
these angles and subsequent terms.

\begin{figure}
\centering
\vspace{9cm}
\includegraphics{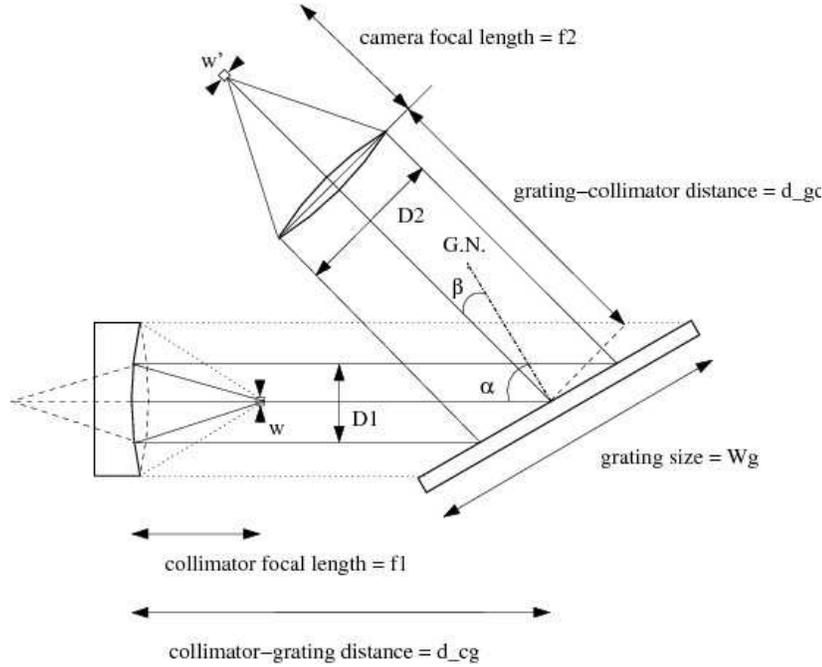}
\caption{Basic spectrograph layout schematic for reflective/refractive
collimator, reflection grating and refractive camera.}
\end{figure}

The system magnification can be broken down into spatial and
anamorphic factors. The physical entrance aperture width, $w$, is
re-imaged onto the detector to a physical width $w'$, demagnified by the
ratio of camera to collimator focal lengths. Hence the spatial width
(perpendicular to dispersion) is given as $w_\theta' = w(f_2/f_1)$.
For non-imaging feeds (i.e., fibers or lenslets), it is advantageous to
pack as much information as possible into a given pixel -- as long as
individual spatial entrance elements can be resolved. This means
cameras must be as fast as possible, relative to their
collimators. For imaging feeds (slits or slicers), the desire to
preserve and sample the spatial information retained in the slit means
the choice must be science-driven.

In the dispersion direction, an additional, anamorphic factor, $r$,
arises due to the fact that grating diffraction implies incident and
diffracted angles need not be the same. Hence incident and diffracted
beam sizes scale as $r = D_1 / D_2 = cos \ \alpha \ / \ cos \
\beta$. This arises because in general $A \times \Omega$ is conserved;
if the beam gets larger, the angles get smaller. Another way to think
of this is in terms of the definition of $r = | d\beta/d\alpha |$, and
ask: For a given d$\alpha$ (angular slit width) what is d$\beta$ such
that d$\lambda = 0$? This result can then be derived from the grating
equation. In any case, $\beta/\alpha>1$ implies magnification, while
$\beta/\alpha<1$ implies demagnification. The re-imaged slit-width in
the spectral dimension is then $w_\lambda' = r w_\theta'$. In Littrow
configurations, important below, $\alpha = \beta =
\delta$ (the latter being the grating blaze angle), and so there is no
anamorphic factor. Since the re-imaged slit-width always degrades the
instrumental spectral resolution, it is always advantageous, in this
sense, to have anamorphic demagnification. However, depending on the
pixel sampling, optical aberrations, and slit size, $w_\lambda'$ may
not be the limiting factor in instrumental resolution. Anamorphic
demagnification also comes at a cost: The camera must be large (larger
than the collimator) to capture all of the light in the expanded
beam. Demagnification never hurts resolution, but the cost should be
weighed against the gains.

The spectral resolution can now be written as $R = \lambda /
d\lambda$, or $R = \lambda (\gamma/r) (f_1/w)$. The term $\gamma/r$
indicates we want large dispersion, but that we can get resolution
also from anamorphic demagnification. The terms $f_1/w$ indicates we
want a {\it long} collimator at fixed camera focal-length, requiring a
field lens or white-pupil design to avoid vignetting.\footnote{A
field-lens, which sits near a focus to avoid introducing power into
the beam, serves to move the spatial pupil to a desirable location in
the system.  This is often the grating, but in in general can be the
location such that the overall system-vignetting is minimized.  A
white pupil design (Baranne 1972, Tull et al. 1995) is one which
re-images a pupil placed on a grating, typically onto a second grating
(e.g., a cross-disperser) or the camera objective.  It is ``white''
because the pupil image location is independent of wavelength even
though the light is dispersed.} Alternatively, we may re-write the
equation as $R = \lambda (\gamma/r) (D_1/\theta D_T)$ noting $\theta$
is the angle on the sky, $d\lambda = w_\lambda' / (dl/d\lambda)$, $w =
f_T \theta$, and $f_1/d_1 = f_T/D_T$, where $f_T/D_T$ refer to the
effective focal-ratio of whatever optics feed the spectrograph, e.g.,
the telescope. The combination of $r$ and $D_1$ indicates we want a
larger collimator and an even larger camera. Using the grating
equation we may write $R = (f_1/w) (sin \ \beta + sin \ \alpha) / cos
\ \alpha$, which, in Littrow configurations reduces to $R = (f_1/w) \
2 \ tan \ \alpha$. In the latter situation it is clear that resolution
can be dispersion-driven by going to large diffraction angles,
$\alpha$, which requires {\it large gratings}.

\subsection{Dispersive Elements}

We distinguish here principally between reflection and transmission
gratings. Transmission gratings yield much more compact spectrograph
geometries. This leads to less vignetting and better performance with
smaller optics.

{\it Reflection gratings} come in three primary varieties: ruled
surface-relief (SR), holographically-etched SR, or volume-phase
holographic. We list the pros and cons of each of these. (i) Ruled SR
gratings have the advantage of control over the groove shape, blaze
and density, which provides good efficiency in higher orders (e.g.,
echelle) at high dispersion. There are existing samples of masters
with replicas giving up to 70\% efficiency, but 50-60\% efficiency is
typical, with 40\% as coatings degrade. Scattered light and ruling
errors can be significant, and existing masters are limited in type
and size. It does not appear to be possible to make larger masters
with high quality at any reasonable cost. (ii) Holographically etched
SR gratings have low scattered light, the capability to achieve high
line-density (hence high dispersion), and large size. However, they
have low efficiency ($<$50\%) because symmetric grooves put equal
power in positive and negative orders. (iii) Volume-phase holographic
gratings can be made to diffract in reflection (Barden et al. 2000),
but have not yet been well-developed for astronomical use.  Reflection
gratings can be coupled to prisms to significantly enhance resolution
via anamorphing (Wynne 1991).

{\it Transmission gratings} are either SR or volume-phase holographic,
and when coupled with prisms are referred to as grisms. (i) SR
transmission gratings and grisms are efficient at small angles and low
line-densities (good for low-resolution spectroscopy), but are
inefficient at large angles and line-densities due to
groove-shadowing. Transmission echelles do exist, but have 30\%
diffraction efficiencies or less. (ii) VPH gratings and grisms are
virtually a panacea. They are efficient over a broad range of
line-densities and angles. Any individual grating is also efficient
over a broad range of angles, (what is known as a broad ``superblaze''
-- see below). Peak efficiencies are as high as 90\%; they are
relatively inexpensive to make, and likewise to customize; and they
can be made to be very large (as larger as your substrate and
recording beam -- now approaching 0.5m). Their only disadvantages is
that they have, to date, been designed for Littrow configurations.

It is worth dwelling somewhat on the theory and subsequent potential
of VPH gratings.  There still remain manufacturing issues of obtaining
good uniformity over large areas (Tamura et al. 2005), but it is
reasonable to be optimistic that refinement of the process will
continue at rapid pace. Application in the near-infrared (NIR) for
cryogenic systems is also promising: CTE mismatch between substrate
and diffracting gelatin, potentially causing delamination, does not
appear to be a concern (W. Brown, private communication, this Winter
School).  Blais-Ouellette et al. (2004) have confirmed that
diffraction efficiency holds up remarkably well at 77K, but that the
effective line-density changes with thermal contraction. We can expect
most grating-fed spectrographs in the future will use VPH gratings
alone or in combination with conventional (e.g., echelle)
gratings. The capabilities of VPH gratings will open up new design
opportunities, many of which will be well suited to 3D spectroscopy.

\subsection{VPH Grating Operation and Design} 

Diffraction arises from modulation of the index of refraction in a
sealed layer of thickness $d$ of dichromated gelatin (the material is
hygroscopic), with mean optical index $n_2$. Typical values for $n_2$
are around 1.43, but the specific value depends sensitively on the
modulation frequency (i.e., the line density $\Lambda$) and
amplitude, $\Delta n_2$, and the specifics of the exposure and
developing process. (Note that it is not currently possible to predict
the precise value of $n_2$ from a manufacturing standpoint.)  The seal
is formed typically by two flat substrates, but this can be
generalized to non-flat surfaces and wedges (i.e., prisms). Because
this layer represents a volume ($d\gg\lambda$), the diffraction
efficiency is modulated by the Bragg condition: $\alpha = \beta$.
These angles are defined here with respect to the plane of the index
modulations. 

The wonder of VPH gratings is the ability to custom design
them. Starting with a science-driven choice of dispersion and
wavelength, the grating equation and dispersion relation given the
Bragg condition uniquely set the line-frequency and angle,
respectively -- for unblazed gratings. The key to high diffraction
efficiency is then to tune the gelatin thickness and index modulation
amplitude such that diffraction efficiency is high in both s and
p-polarizations (the s-polarization electric vector is perpendicular
to the fringes). This can be done by brute force via rigorous coupled
wave calculations, or by noting that in the so-called ``Kogelnik
limit'' the diffraction efficiencies are periodic in these quantities
(Barden et al. 2000; Baldry et al. 2004). The two polarizations have
different periodicities, i.e., VPH gratings are in general highly
polarizing, so the trick is finding the ($d,\Delta n_2$)-combination
that phases one pair of s and p efficiency-peaks.  Thinner gel layers
yield broader band-width over which the diffraction-efficiency is high
-- relative to the efficiency at the Bragg condition. The thinner the
layer, the larger the index modulation required to keep the efficiency
high in an absolute sense. Modulations above 0.1 are very difficult to
achieve, and more typical values are in the range of 0.04 to 0.07; gel
layers are in the range of a few to a few 10's of microns. In
practice, because there is limited manufacturing control over the
index modulation and effective depths of the gelatin exposure,
gratings requiring very precise values in these parameters will be
difficult to make, and have large inhomogeneities. Our experience is
that it is useful to understand how wavelength and resolution
requirements can be relaxed to locate more robust design-parameters.

\subsubsection{Blazed VPH Gratings} 

Nominally the fringe plane is parallel to the substrate normal
(indicated by the angle $\phi=0$). This yields an unblazed
transmission grating. Essentially all astronomical VPH gratings in use
are made this way. There is concern that tilted fringes will curve
with the shrinkage of the gelatin during development (Rallison \&
Schicker 1992), but this concern has not been fully explored. By
tilting the fringes (this is done simply by tilting the substrate
during exposure in the hologram), one can enter several different
interesting regimes, as illustrated concisely by Barden et al. (2000;
see their Figure 1): small $|\phi|$ yields blazed reflection gratings,
$\phi = 90$ deg produces unblazed reflection gratings, and large $|\phi| $
blazes the reflection gratings. ``Large'' and ``small'' depend on the
angle of incidence, as illustrated below. The sign convention is such
that positive $\phi$ decreases the effective incidence angle.  The
incident and reflected angles in the gelatin, $\alpha_2$ and
$\beta_2$, are related by $\alpha_2 = \beta_2 + 2\phi$, with
$\alpha_2-\phi$ being the effective diffraction angle. The grating
equation, when combined with the Bragg condition yields: $m \
\lambda_b \ = \ 2 \ n_2 \ \Lambda \ sin \ (\alpha_2-\phi)$, where
$\lambda_b$ is the Bragg wavelength, and $\Lambda = \Lambda_g \ cos \
\phi$ is the fringe spacing perpendicular to the fringes.  We use
Baldry et al.'s (2004) nomenclature; their Figure 1 is an instructive 
reference for this discussion.

\begin{figure}
\centering
\vspace{9cm}
\includegraphics{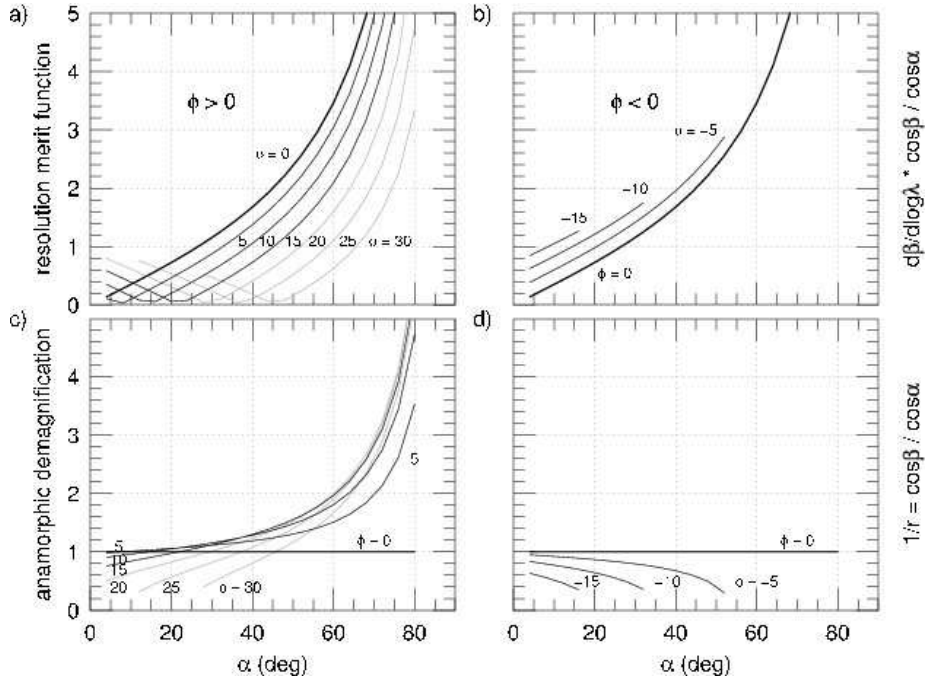}
\caption{Resolution merit function and anamorphic factor for blazed
VPH gratings with mean gel index n$_2 = 1.43$. Typical
SR gratings have $1.05 < 1/r < 1.2$.}
\end{figure}

Baldry et al. work out the case for no fringe tilt with flat or wedged
substrates.  Here we give the case of flat substrates but arbitrary
$\phi$. Burgh et al. (2007) extend this to include arbitrary fringe
tilt. The relevant angles with respect to the grating normal can be
found with these equations in terms of the physical grating
properties:
$$sin \ \alpha \ = \ n_2 \ sin \ \alpha_2, \ \ {\rm and} \ \ \ sin \
\beta \ = \ n_2 \ sin[\ sin^{-1} (\frac{sin \ \alpha}{n_2}) - 2\phi \
].$$ The anamorphic factor and dispersion are still defined in terms
of $\alpha$ and $\beta$ as given in \S 1.2.1. With the interrelation
of these angles as given above, it is easy to show the logarithmic
angular dispersion at the Bragg wavelength is:
$$d\beta/dlog\lambda = 2 \ n_2 \ cos \ \phi \ sin[\ sin^{-1}
(\frac{sin \ \alpha}{n_2})-\phi] \ / \ cos \ \beta.$$

To understand the potential advantages of blazed transmission
gratings, we define a resolution merit function as $\frac{1}{r} \
d\beta/dlog\lambda,$ i.e., the product of the logarithmic angular
dispersion and the anamorphic factor. With this function we can
explore, in relative terms, if tilting the fringes yields resolution
gains. Figure 1.6 shows the anamorphic factor and the resolution merit
function versus grating incidence angle for positive and negative
fringe-tilts. Negative fringe tilts give a small amount of increased
resolution at a given $\alpha$ by significantly increasing
dispersion, which over-comes an increase in the anamorphic {\it
mag}nification. This means the detector is less efficiently
used. Negative fringe tilts also limit the usable range of $\alpha$
for which $\beta < 90$ deg (transmission), and hence the maximum
achievable resolution in transmission that can be achieved
is lowered with negative fringe tilts.

With positive fringe tilts, the anamorphic demagnification increases
strongly at large incidence angles, although there is little gain in
going to $\phi > 15$ deg. Note that the demagnification becomes $<1$
(i.e., magnification) roughly when $\alpha \sim 1.5 \ \phi$.  This is
when the effective diffraction angle ($\alpha_2-\phi$), changes sign
with respect to the tilted fringes (the grating remains in
transmission). The overall resolution decreases with increased
positive fringe tilt, but the decrease is modest for small tilt
angles. Given the large increase in anamorphic demagnification
relative to the modest loss in resolution, for small tilt angles there
is a definite gain in information: A +5 deg tilt gives a 12\% loss in
the resolution merit function at $\alpha=60$, but a 51\% gain in the
anamorphic demagnification. With suitably good optics and detector
sampling the demagnified image, this equates directly into an increase
in the number of independent spectral resolution elements, replete
with a 72\% increase in spectral coverage.  The loss in resolution can
easily be made up by slightly increasing $\alpha$ (in this case, from
60 to 63 deg) and modulating $\Lambda$ in the grating design to tune
the wavelength. Instruments with blazed, high-angle VPH gratings with
tilts of $5< \phi < 15$ deg will allow for the high resolution needed to
work between sky-lines, while efficiently packing spectral elements
onto the detector.  This is critical in the context of 3D
spectroscopy, where room must also be made for copious spatial
elements.

% alpha = 60
% phi      0     5      10    15
% merit    3.464 3.043  2.581 2.094
% 1/r      1.000 1.511  1.811 1.967
% alam     3.464 2.014  1.426 1.065 
% a(m0)    60.0  63.0   66.0  69.5   
% where a(m0) is the alpha to achieve the alpha = 60 phi = 0 merit.

\begin{figure}
\centering
\vspace{8cm}
\includegraphics{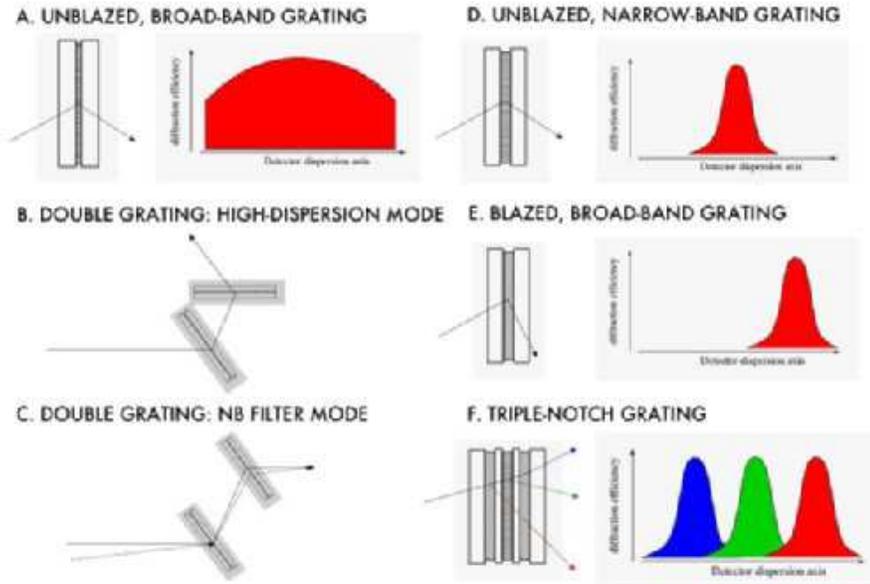}
\caption{Novel grating modes. A. Conventional broad-band application
now becoming a staple of modern spectrographs.  B. Double-grating
geometry yielding a net dispersion of $\sim$ the sum of the two
individual grating dispersions (gratings are not necessarily
identical, but angles must be adjusted accordingly). C. Double-grating
geometry yielding a narrow-band filter with field-dependent band-pass
given by the Bragg condition (gratings are identical) D.-F.  Narrow
band-pass gratings unblazed (D), blazed (E), and combined (F) to form
a {\it notch} grating. Other modes are discussed in the text. Panels A, D-F
show both the grating configuration as well as a cartoon-sketch of the
diffraction efficiency as a function of location of the detector
dispersion axis, labeled for the mean wavelength regime of the
diffraction band-pass.}
\end{figure}

\subsubsection{Unusual VPH Grating Modes} 

In addition to tilted fringes, VPH gratings pose opportunities for a
number of novel modes well suited to 3D spectroscopy. Figure 1.7
illustrates some of these. With very high diffraction efficiency it is
now reasonable to consider combining gratings to augment the
dispersion, and hence resolution. If the two gratings are kept
parallel but offset along the diffraction angle, they can serve as
(tunable) narrow-band filters -- an alternative to etalons (e.g.,
Blais-Ouellette et al. 2006). Barden et al. (2000) have explored using
multiple gelatin layers with different line-frequencies to select
H$\alpha$ and H$\beta$ in separate band-passes. By slightly rotating
one set of lines, sufficient cross-dispersion is added to space the
two spectra -- one above the other -- on the detector. This is well
suited for spectrographs fed with widely spaced fibers or slitlets
(i.e., an under-filled, conventional long-slit spectrograph), and
represents an interesting trade-off in wavelength and spatial
multiplex. At sufficiently high dispersion (and hence limited
band-pass), the number of layers could be increased to mimic a
multi-order echelle.\footnote{A true cross-dispersed echelle-like
grating would work, in principle, with two layers, rotated by 90
degrees. VPH gratings have not yet been made with high efficiency in
multiple orders, but see Barden et al. (2000) for measurements up to
order 5.}  The advantage of this approach is in resolution and
wavelength coverage.

An alternative approach is something we refer to as {\it ``notch''}
gratings. Here, we take advantage of the relative ease (from a
manufacturing stand-point) of achieving a narrow band-pass, and
combine gel layers tuned to different, non-over-lapping wavelength
band-passes at a given incidence angle (e.g., by changing the line
frequency). By also tuning the fringes with with modest tilts, each
band pass can be centered on a different, non-overlapping portion of
the detector. Band-passes will have to be carefully crafted by tuning
grating parameters to avoid parasitic contamination in the other
bands. The figure illustrates positive and negative tilts, but the
tilts could be arranged to all be positive to take advantage of the
anamorphic factors described above. This offers another way to slice
the data cube -- one which allows for {\it sparse} spectral sampling
of key spectral diagnostics over a {\it broad} wavelength range (e.g.,
[OII]$\lambda$3727, H$\beta$+[OIII]$\lambda\lambda$4959,5007, and
H$\alpha$) at {\it high} dispersions, with ample room left over on the
detector for significant spatial multiplex.

\subsection{Summary of Implications for 3D Spectrograph Design}

The most compact spectrograph designs yield the highest-efficiency,
wide-field systems needed to grapple with attaining large angular
coverage for 3D spectroscopy. To also obtain high-enough spectral
resolution to work between the atmospheric air-glow often requires
significant dispersive power and anamorphic demagnification.  Large
anamorphic demagnification, while not free (larger camera optics are
required), is well-suited to packing information onto the detector.
This is particularly important in 3D applications where spatial
multiplex is at a premium. VPH transmission gratings are clearly
preferred because they lend themselves to compact spectrograph
geometry and provide high diffraction efficiency. We have shown they
can, in principle, also yield large anamorphic demagnification. With
high-angle, double, and blazed VPH gratings, echelle-like resolutions
can be achieved at unprecedented efficiency (75-90\% in diffraction
alone). Unusual modes to produce tunable narrow-band filters and notch
gratings also open up the possibility for well-targeted sparse,
spectral sampling.

\subsection{Coupling Formats and Methods: Overview}

The essence of the 3D spectrometer lies in the coupling of the
telescope focal plane to the spectrograph. We review the four
principal methods: (i) direct fibers, (ii) fibers + lenslets, (iii)
image-slicers, and (iv) lenslet arrays, or pupil-imaging
spectroscopy. A nice, well-illustrated overview can be found in
Allington-Smith \& Content (1998); additional discussion of the merits
and demerits of different approaches can be found in Alighieri et
al. (2005). Here we also make an evaluation.  We discuss a fifth mode
not seen in the literature, which we refer to as (v) ``filtered
multi-slits.'' Many spectrographs either have, or could easily be
modified to have, this capability. We also describe (vi) multi-object
configurations -- a mode which will undoubtedly become more common in
the future.

Throughout this discussion, we distinguish between near-field versus
far-field effects.  The near-field refers to the light distribution at
the focal surface, e.g., fiber ends, and what is re-imaged ultimately
onto the CCD. The far-field refers to the ray-bundle distribution,
i.e., the cross-section intensity profile of the spectrograph beam
significantly away from the focal surface. Different coupling methods
offer the ability to remap near- and far-field light-bundle
distributions, which can have advantages and dis-advantages.

\subsection{Direct Fiber Coupling}

The simplest and oldest of methods consists of a glued bundle of bare
fibers mapping the telescope to spectrograph focal surfaces. With
properly doped, AR-coated fibers throughput can be at or above 95\%,
which can be compared to 92\% reflectivity off of one freshly coated
aluminum surface. These have the distinct advantage of low cost and
high throughput. As with all fiber-based coupling, there is a high
degree of flexibility in terms of reformatting the telescope to
spectrograph focal-surfaces (for example, it is easy to mix sky and
object fibers along slit), and the feeds can be integrated into
existing long-slit, multi-object spectrographs.  However, bare fiber
IFUs are not truly integral, and do not achieve higher than 60-65\%
fill-factors (see Oliveria et al. 2005 on the deleterious effects
of buffer-stripping of small fibers). This coupling is perhaps the
most cost-effective mode for cases where near-integral sampling is
satisfactory, and preservation of spatial information is not at a
premium.

{\it Information loss and stability gain with fibers:} Focal Ratio
Degradation (FRD) and azimuthal scrambling represent information loss
(an entropy increase). FRD specifically results in a faster output
f-ratio (Ramsey 1988). This has an impact on spectrograph design or
performance since either the system will be lossy (output cone
over-fills optics), or the spectrograph has to be designed for the
proper feed f-ratio. PMAS (Roth et al. 2005) is an excellent example
of how to properly design a spectrograph to handle fast fiber-output
beams.  The existing WIYN Bench spectrograph is a good example of how
not to do it. In fact, it's so bad we rebuilt it (Bershady et
al. 2008); we were able to recapture 60\% of the light (over a factor
of 2 in throughput) with no loss of spectral resolution in the
highest-resolution modes.

Azimuthal scrambling can help and hurt. While scrambling destroys
image information, it symmetrizes the output beam, ameliorating, to
some extent, the effect of a changing telescope pupil on HET or
SALT-like telescopes by homogenizing the ray bundle.  Thus, the
contribution of spectrograph optical aberrations to the final spectral
image is more stable. (This is a far-field effect.)

Radial and azimuthal scrambling together homogenize near-field
illumination, e.g., the seeing-dependent slit function is decreased.
Radial scrambling and FRD are one and the same ({\it cf.} Ramsey 1988
and Barden et al. 1993), so that one trades information loss for
stability (similar to the trade of precision for accuracy).  In
practice, fiber-input beam-speeds of f/3 (PMAS) to f/4.5 (HET and
SALT) are desirable. However, with fast input/output f-ratios this
limits possible spectrograph demagnification since it is expensive to
build faster than f/2 for large cameras.

{\it Telecentricity.} Because azimuthal scrambling symmetrizes a beam,
if the input light-cone is mis-aligned with the fiber axis, the output
beam (f-ratio) is faster.  This is not FRD. To avoid this effect,
fiber telecentric alignments of under a degree are needed even for
f-ratios as fast as 4-6 (Bershady 2004, Wynne \& Worswick 1989).

\begin{figure}
\centering
\vspace{7.5cm}
\includegraphics{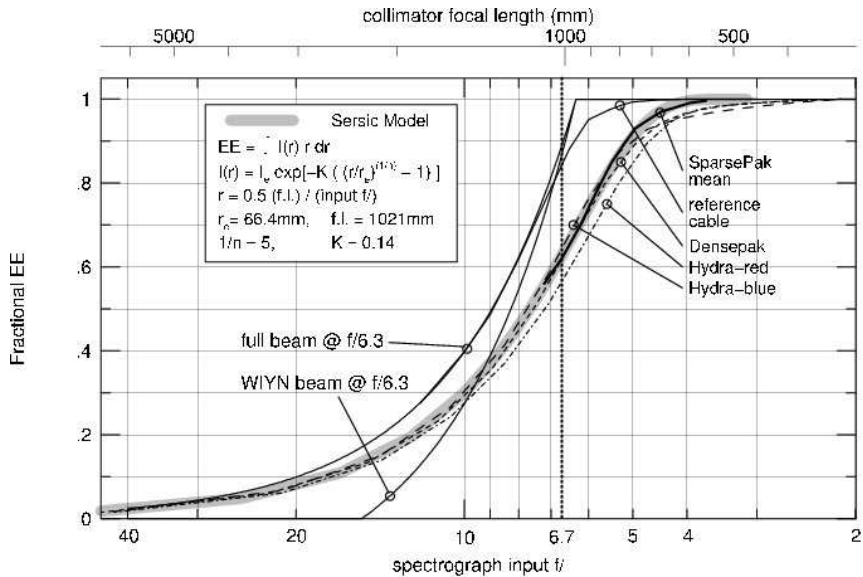}
\caption{Output fiber irradiance (encircled-energy versus beam f-ratio)
for fiber cables on the WIYN Bench Spectrograph. The input beam
profile is an unappodized f/6.3 beam with an f/17 central obscuration
(labeled). Output beam profiles are faster, due to FRD, and are
well-fit by a Sersic model of index 1/n = 5 (S. Crawford, private
communication).}
\end{figure}

{\it Causes of FRD.}  Excessive FRD in fibers is due to
stress. Hectospec (Fabricant et al. 2005) embodies an excellent
example of how to properly treat fibers and fiber cabling (Fabricant
et al. 1998; see also Avila et al. 2003 in the context of FLAMES on
VLT). Fiber termination and polishing can also induce stress.
Bershady et al. (2004) discuss some other IFU-related issues in terms
of buffering. However, even for perfectly handled fibers, there is
internal scattering - the cause of which has long been a
debate. Nelson et al. (1988) suggested a combination of (a) Rayleigh
scattering (variation in fiber refractive index); (b) Mie scattering
(fiber inhomogeneities comparable to the wavelength); (c)
stimulated Raman and Brillouin scattering (not relevant at low signal
level in astronomical applications); and (d) micro-bending.
Micro-bending seems like a good culprit; it is the unsubstantiated
favorite in the literature. Micro-bending models predict a
wavelength-dependent FRD. While Carrasco \& Parry (1994) tentatively
see such an effect, neither Schmoll et al. (2003) or Bershady et
al. (2004) confirm the result.  However, these studies use different
measurements methods. More work is required to understand the physical
cause(s) of FRD, and with this understanding, perhaps, reduce the
amplitude of the effect. We find FRD produces an output fiber beam
profile which can be well-modeled by a Sersic function (Figure 1.8;
S. Crawford, private communication). This either says something about
the scattering model or how seriously to take physical interpretations
of Seric-law profiles of galaxies{\it !}

\begin{figure}
\centering
\vspace{2cm}
\includegraphics{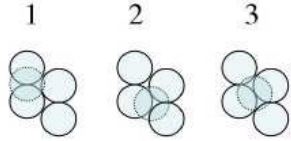}
\caption{Critical sampling with densely-packed fibers.}
\end{figure}

{\it Quality versus quantity:} Fibers offer the opportunity of easily
trading quality for quantity in terms of packing the spectrograph
slit. Scattered light within the spectrograph, combined with fiber
azimuthal-scrambling means spatial information in the telescope focal
plane is coupled to all adjacent fibers in the slit. Closely packing
fibers in the slit can make clean spectral extraction difficult.  The
WIYN Bench spectrograph is a good example where the amplitude of
scattered light is low, fiber separation is large and ghosting is
negligible. This spectrograph and feeds are optimized for clean
extraction with little cross-talk (1\% cross-talk in visible in
optimum $S/N$ aperture, degrading to 10\% in the NIR).  Information
packing in the spatial dimension is modest due to fiber separation,
while information packing in the spectral dimension is high due to
large anamorphic factors.  Other systems have significant spectral
overlap. For example, staggered slits, where fibers are separated by
only their active diameter (COHSI; Kenworthy et al. 1998) make it
difficult to extract a clean spectrum and optimize $S/N$ at the same
time, but the spatial multiplex is increased. There is no one right
answer, but definitely a decision worthy of a science-based
consideration.

{\it Image reconstruction and registration.} Even without lenslets,
densely sampled fibers provide excellent image reconstruction on
spatial scales of order the fiber diameter. One can achieve the
theoretical sampling-limit with a 3-position pattern of
half-fiber-diameter dithers (Figure 1.9; {\it cf.} Koo et al. 1994 in
the context of under-sampled HST/WFPC-2 data). Even with sparse
sampling, registration of the spectral data-cube with broad-band
images can be achieved to 10\% of the fiber diameter by
cross-correlating the spectral continuum with respect to broad-band
images or integrated radial light profiles (Bershady et al. 2005).
Kelz et al. (2006) show how well it is possible to reproduce the
continuum image of UGC 463 using the PPak fiber bundle -- without any
sub-sampling.

\begin{figure}
\centering
\vspace{7.5cm}
\includegraphics{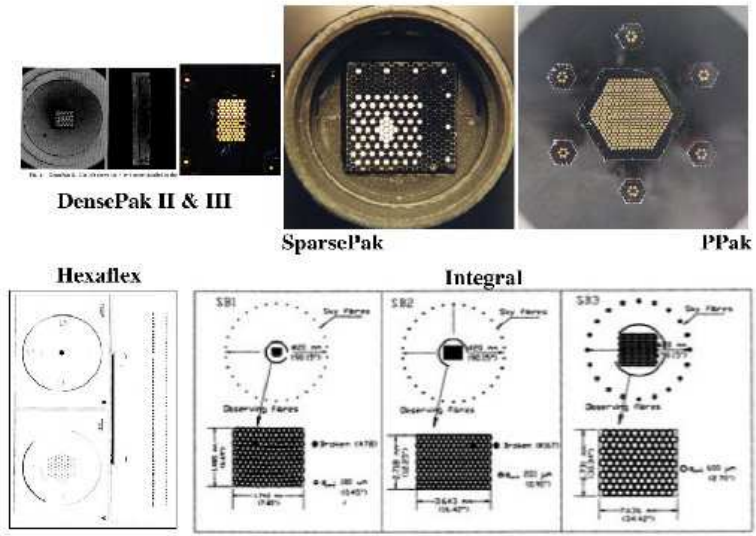}
\caption{Direct-fiber IFUs on optical spectrographs. The top row shows
the legacy started by S. Barden with DensePak-1 and DensePak, leading
to SparsePak, PPaK on the KPNO 4m, WIYN, and Calar Alto, respectively;
the bottom row shows Hexaflex and Integral on WHT with their
multiple, selectable bundles and ample sky-fibers.}
\end{figure}

{\it Summary of instruments.} Some of the first IFUs were on the KPNO
4m RC spectrograph: DensePak-1 followed by DensePak-2 (Barden \& Wade
1988; see also Guerrin \& Felenbok 1988 for other early IFUs). The
last incarnation (Barden et al. 1998) was on WIYN. Conceptually, these
instruments spawned SparsePak (WIYN; Bershady et al. 2004) and PPak
(PMAS, Calar Alto, Verheijen et al. 2004; Kelz et al. 2006). A
more-versatile single instrument-suite, built for the WHT, is INTEGRAL
(WYFFOS), which offers several plate-scales and formats (Arribas et
al. 1998), and a sophisticated and well thought-out mapping between
telescope and spectrograph focal planes. These are all shown in Figure
1.10. GOHSS is one case of a NIR (0.9-1.8$\mu$m) application
(Lorenzetti et al. 2003). VIRUS (Hill et al. 2004) and APOGEE (Allende
Prieto et al. 2008) are the only planned future instruments.

\begin{table}
\caption{Direct Fiber-Coupled Integral Field Instruments}
\tiny
\begin{tabular}{llllllllll}\hline \hline 
Instrument &  Tel. & D$_T$ & $\Omega$ & d$\Omega$ &
N$_\Omega$ & $\Delta\lambda/\lambda$ & R & N$_R$ & $\epsilon$ \\
& & (m) & \multicolumn{2}{c}{(arcsec$^2$)} &  &  &  &  &  \\ \hline
\multicolumn{10}{c}{Existing Optical Instruments} \\ \hline 
DensePak & WIYN & 3.5 & 564 & 6.2 & 91 & 1.02 & 1000 & 1024 & 0.04\\
         &      & 3.5 & 564 & 6.2 & 91 & 0.07 & 13750 & 1024 & 0.04\\
         &      & 3.5 & 564 & 6.2 & 91 & 0.04 & 24000 & 1024 & 0.04\\
         &      & 3.5 & 119 & 1.3 & 91 & 1.02 & 1000 & 1024 & 0.04\\
         &      & 3.5 & 119 & 1.3 & 91 & 0.07 & 13500 & 1024 & 0.04\\
         &      & 3.5 & 119 & 1.3 & 91 & 0.04 & 24000 & 1024 & 0.04\\
SparsePak& WIYN & 3.5 & 1417 & 17.3 & 82 & 1.02 & 800 & 819 & 0.07\\
         &      & 3.5 & 1417 & 17.3 & 82 & 0.07 & 11000 & 819 & 0.07\\
         &      & 3.5 & 1417 & 17.3 & 82 & 0.03 & 24000 & 819 & 0.07\\
PPak     & CA   & 3.5 & 2070 & 5.64 & 367 & 0.15 & 7800 & 1183 & 0.15\\
INTEGRAL & WHT  & 4.2 & 32.6 & 0.159 & 205 & 0.22 & 2350 & 515 & $\cdots$\\
         &      & 4.2 & 32.6 & 0.159 & 205 & 0.94 & 550 & 515 & $\cdots$\\
         &      & 4.2 & 139.3 & 0.64 & 219 & 0.22 & 2350 & 515 & $\cdots$\\
         &      & 4.2 & 139.3 & 0.64 & 219 & 0.94 & 550 & 515 & $\cdots$\\
         &      & 4.2 & 773 & 5.73 & 135 & 0.07 & 2350 & 300 & $\cdots$\\
         &      & 4.2 & 773 & 5.73 & 135 & 0.90 & 550 & 300 & $\cdots$\\ \hline
\multicolumn{10}{c}{Future Optical Instruments} \\ \hline 
VIRUS    & HET & 9.2 & 32604 & 1.0 & 32604 & 0.505 & 811. & 410 & 0.16\\ \hline
\multicolumn{10}{c}{Existing Near Infrared Instruments} \\ \hline 
GOHSS    & TNG & 3.6 & 44.2 & 1.77 & 25 & 0.12 & 4380. & 512 & 0.13\\ \hline
\multicolumn{10}{c}{Future Near-Infrared Instruments} \\ \hline 
\multicolumn{10}{c}{} \\ [-0.08in] \hline \hline
\end{tabular}
\end{table}

\subsection{Fiber + Lenslet Coupling}

The basic concept of lenslet coupling to fibers is again, as with bare
fibers, to remap a 2D area in the telescope focal-surface to a 1D slit
at the spectrograph input focal surface. The key difference is in the
fore-optics, which consists of a focal expander and lenslet array;
these feed the fiber bundle. The focal expander serves to matches
to the scale of the lenslet array. Allington-Smith \& Content (1998)
and Ren \& Allington-Smith (2002) present some technical discussion
and illustration of the method. Each micro-lens in the array then
forms a pupil image on the fiber input face.  The pupil image is
suitably smaller than the lenslet to allow the fibers to be packed
behind the integral lenslet array.  This reduction speeds up the input
beam ($A\times\Omega$ is conserved). Given the previous discussion
concerning FRD, this can be advantageous to minimize entropy increase.

At the output stage, the option exists to reform the (now azimuthally
scrambled) slit-image with an output micro-lens linear array, or to
use bare fibers. Without lenslets, the input f-ratio to the
spectrograph will be faster, which means there is less possibility for
geometric demagnification via a substantially faster camera. 
In this case the spectrograph also reimages the fiber-scrambled
telescope pupil: the image varies with telescope illumination, while
the ray-bundle distribution (far-field) varies with the telescope
image.

The positive attributes of lenslet-fed fiber arrays are: (i) improved
filling factors to near unity; and (ii) control of input and output
fiber f-ratio. The latter permits effective coupling of a slow telescope
f-ratio to fiber input at a fast, non-lossy beam speed, and likewise,
permits effective coupling of fiber output to spectrograph.  The
negative attributes of this coupling method include (iii) increased
scattered light (from the lenslet array); (iv) lower throughput (due
to surface-reflection, scattering, and misalignment).  For example,
typical lenslet + fiber units yield only 60-70\% throughput
(Allington-Smith et al. 2002).  When there is a science premium on
truly integral field sampling, the above two factors don't out-weigh
the filling factor improvements.  Finally, there is the more subtle
effect of whether or not to use output lenslets.  Aside from the
matter of f-ratio coupling, there is the issue of whether swapping the
near- and far-field patterns is desirable for controlling systematics
in the spectral image. It amounts to assessing whether the
spectrograph is ``seeing-limited'', i.e., limited by spatial changes in the
light distribution within the slit image formed by the fiber and
lenslet, or aberration limited?

Prime examples of optical instruments on 8m-class telescopes include
VIMOS (Le Fevre et al. 2003), GMOS (Gemini-N,S, Allington-Smith et
al. 2002), and FLAMES/GIRAFFE in ARGUS or multi-object IFU modes
(Avila et al. 2003)\footnote{See also
www.eso.org/instruments/flames/inst/Giraffe.html.}.  Typical
characteristics of these devices is fine spatial sampling (well under
an arcsec) and modest spectral resolution. ARGUS is an exception,
achieving resolutions as high 39,000.  It's multi-object mode is also
unique -- and powerful (see later discussion).  On 4m-6m class
telescopes there are PMAS (Roth et al. 2005), Spiral+AAOmega (Saunders
et al. 2004, Kenworthy et al. 2001), MPFS (Afanasiev et
al. 1990)\footnote{See also
www.sao.ru/hq/lsfvo/devices/mpfs/mpfs\_main.html.}, and IMACS-IFU
(Schmoll et al. 2004).\footnote{See also
www.lco.cl/lco/magellan/instruments/IMACS/.} Compared to most
direct-fiber IFUs on comparable telescope, these instruments also have
finer spatial sampling.

\begin{table}
\caption{Fiber+Lenslet Coupled Integral Field Instruments}
\tiny
\begin{tabular}{llllllllll}\hline \hline
Instrument & Tel. & D$_T$ & $\Omega$ & d$\Omega$ &
N$_\Omega$ & $\Delta\lambda/\lambda$ & R & N$_R$ & $\epsilon$ \\
& Method & (m) & \multicolumn{2}{c}{(arcsec$^2$)} & &  &  &  &  \\ \hline
\multicolumn{10}{c}{Existing Optical Instruments} \\ \hline 
PMAS & Calar Alto & 3.5 & 64.  & 0.5  & 256 & 0.11 & 9400 & 1000 & 0.15\\
     &            & 3.5 & 64.  & 0.5  & 256 & 0.52 & 1930 & 1000 & 0.15\\
     &            & 3.5 & 144. & 0.75 & 256 & 0.11 & 9400 & 1000 & 0.15\\
     &            & 3.5 & 144. & 0.75 & 256 & 0.52 & 1930 & 1000 & 0.15\\
     &            & 3.5 & 256. & 1.0  & 256 & 0.11 & 9400 & 1000 & 0.15\\
     &            & 3.5 & 256. & 1.0  & 256 & 0.52 & 1930 & 1000 & 0.15\\
SPIRAL& AAT       & 3.9 & 251. & 0.49 & 512 & 0.29 & 1700 & 495 & 0.25 \\
     &            & 3.9 & 251. & 0.49 & 512 & 0.07 & 7500 & 495 & 0.25 \\
MPFS & SAO        & 6.0 & 256. & 1.0  & 256 & 0.12 & 8800 & 1024 & 0.045\\
     &            & 6.0 & 64.  & 0.25 & 256 & 0.47 & 2200 & 1024 & 0.045\\
IMACS-IFU & Magellan &6.5& 62.0& 0.031& 2000& 0.61 & 2500  & 4096 & 0.19 \\
     &            & 6.5 & 37.7 & 0.031& 1200& 0.31 & 7500  & 2340 & 0.17 \\
GMOS & Gemini & 8.0 & 49.6 & 0.04 & 1500 & 0.21 & 3450 & 730. & $\cdots$\\
     &            & 8.0 & 49.6 & 0.04 & 1500 & 0.32 & 2300 & 730 & $\cdots$\\
     &            & 8.0 & 49.6 & 0.04 & 1500 & 0.82 & 890 & 730 & $\cdots$\\
     &            & 8.0 & 24.8 & 0.04 & 750 & 0.42 & 3450 & 1460 & $\cdots$\\
     &            & 8.0 & 49.6 & 0.04 & 1500 & 0.64 & 2300 & 1460 & $\cdots$\\
     &            & 8.0 & 49.6 & 0.04 & 1500 & 1.00 & 890 & 1460 & $\cdots$\\ 
VIMOS & VLT       & 8.0 & 2916. & 0.45 & 6400 & 0.6 & 250 & 150 & $\cdots$\\
     &            & 8.0 & 698. & 0.11 & 6400 & 0.6 & 250 & 150 & $\cdots$\\
     &            & 8.0 & 729. & 0.45 & 1600 & 0.2 & 2500 & 500 & $\cdots$\\
     &            & 8.0 & 174.5 & 0.11 & 1600 & 0.2 & 2500 & 500 & $\cdots$\\
ARGUS/IFU & VLT   & 8.0 & 83.9 & 0.27 & 315 & 0.105 & 11000 & 1155 & $\cdots$\\
     &            & 8.0 & 83.9 & 0.27 & 315 & 0.042 & 39000 & 1625 & $\cdots$\\
ARGUS & VLT & 8.0 & 27.7 & 0.09 & 315 & 0.105 & 11000. & 1155 & $\cdots$\\
     &            & 8.0 & 27.7 & 0.09 & 315 & 0.042 & 39000. & 1625 & $\cdots$\\ \hline
\multicolumn{10}{c}{Future Optical Instruments} \\ \hline 
\multicolumn{10}{c}{Existing Near-Infrared Instruments} \\ \hline 
COHSI & UKIRT     & 3.8 & 8.5  & 0.85 & 100 & 0.26 & 500. & 128 & $\cdots$\\
SMIRFS & UKIRT    & 3.8 & 24.2 & 0.34 & 72 & 0.023 & 5500. & 128 & $\cdots$\\
CIRPASS & Gemini  & 8.0 & 54.5 & 0.13 & 490 & 0.41 & 2500. & 1024 & $\cdots$\\
     &            & 8.0 & 54.5 & 0.13 & 490 & 0.085 & 12000. & 1024 & $\cdots$\\
     &            & 8.0 & 27.0 & 0.06 & 490 & 0.41 & 2500. & 1024 & $\cdots$\\
     &            & 8.0 & 27.0 & 0.06 & 490 & 0.085 & 12000. & 1024 & $\cdots$\\ \hline
\multicolumn{10}{c}{Future Near-Infrared Instruments} \\ \hline 
\multicolumn{10}{c}{} \\ [-0.08in] \hline \hline
\end{tabular}
\end{table}

NIR instruments include SMIRFS (Haynes et al. 1999), and COHSI, which
is a precursor - in some regards - to CIRPASS (Parry et al. 2004).  An
interesting application of flared fibers is discussed by Thatte et
al. (2000) for cryogenic systems.

A summary of existing and future optical and NIR lenslet + fiber
coupled IFU spectrographs are listed in Table 2. While it may seem
surprising that no future instruments appear to be planned, we will
discuss one possible instrument for the 30m Telescope (TMT) below.

\subsection{Slicer Coupling}

Image-slicers have been around for a long time, primarily serving the
high-resolution community, e.g., to slice a large fiber into a thin,
relatively short slit to feed cross-dispersed echelle's (see Tull et
al. 1995 for one recent example). Extending the concept into a 3D mode
follows the same basic notion, which can be thought of as deflecting
slices of the telescope image plane both along and perpendicular to
the slice through a pair of reflections. These reflections have power
to reform the focal-plane image. Given the deflections, the
slices are re-aligned end-to-end as in a long-slit, which then feeds
a conventional spectrograph. 

The latest incarnation is the so-called ``Advanced Image Slicer''
(AIS) concept -- a 3-element system, introduced and nicely illustrated
by Allington-Smith et al. (2004). In short, the slicer mirrors at the
telescope focal plane divide it into strips, and have power to place
the telescope pupil on the next slicer element. This is desirable to
keep these elements small and the slicer compact. The second element
is an array of pupil mirrors (one per slice), which reformat the
slices into a pseudo-slit, where they form an image of the sky. A
tertiary field lens (a lenslet for each slice) control the location of
the pupil stop in the spectrograph.  This is critical for efficient
use of the spectrograph.  All-mirror designs exist for the NIR
(FISICA, Eikenberry 2004b), taking advantage of lower scattering at
longer wavelengths to machine monolithic elements.  Catadioptric
designs exist for the optical (MUSE, Henault et al. 2004). Here the
pupil lenses replace pupil mirrors, which aids the geometric layout of
the spectrograph system.

The salient features of image slicers are (i) they are the only IFU
mode to preserve all spatial information. All other coupling modes
destroy spatial information within the sampling element, either by
fiber scrambling or pupil-imaging (below).  (ii) Image slicers are
also the most compact at reformatting the focal plane onto the
detector.  (iii) They can be used in cryogenic systems and at long
wavelengths where fibers don't transmit (although lenslet arrays also
accomplish this -- see next section). There are some disadvantages,
including (iv) scattered-light from the slicing mirrors
(diamond-turned optics can't be used in the optical), and (v) a lack
of reformatting freedom. The latter is perhaps less of a concern given
that the image is being preserved. However, for possible multi-object
modes, particular attention must be payed to the design of the
required relay optics to avoid efficiency losses.

\begin{table}
\caption{Slicer Coupled Integral Field Instruments}
\tiny
\begin{tabular}{llllllllll}\hline \hline
Instrument & Tel. & D$_T$ & $\Omega$ & d$\Omega$ &
N$_\Omega$ & $\Delta\lambda/\lambda$ & R & N$_R$ & $\epsilon$ \\
 &  & (m) & \multicolumn{2}{c}{(arcsec$^2$)} & &  &  &  &  \\ \hline
\multicolumn{10}{c}{Existing Optical Instruments} \\ \hline 
ESI$^a$     & Keck    & 10.0 & 22.8 & 1.28 & 18 & 0.95 & 3500  & 3325  & 0.14 \\ 
            &         & 10.0 & 15.0 & 0.56 & 27 & 0.95 & 5200  & 4950  & 0.14 \\ 
            &         & 10.0 & 10.0 & 0.25 & 40 & 0.95 & 7800  & 7410  & 0.14 \\ 
            &         & 10.0 &  8.4 & 0.09 & 93 & 0.95 & 13000 & 12350 & 0.14 \\ \hline
\multicolumn{10}{c}{Future Optical Instruments} \\ \hline 
WiFeS       & ANU     & 2.3 & 775. & 1. & 775 & 1.03 & 3000 & 3090 & $\cdots$\\
            &         & 2.3 & 775. & 1. & 775 & 0.44 & 7000 & 3090 & $\cdots$\\ 
% IMACS/GISMO & Magellan & 6.5 & $\cdots$ & $\cdots$ & $\cdots$ & $\cdots$ & $\cdots$ & $\cdots$ & $\cdots$ \\ 
MUSE$^a$    & VLT     & 8.0 & 3600 & 0.04 & 9e4 & 0.67 & 3000 & 2000 & 0.24\\\hline
\multicolumn{10}{c}{Existing Near-Infrared Instruments} \\ \hline 
UIST        & UKIRT   & 3.8 & 19.8 & 0.06 & 344 & 0.15 & 3500 & 512 & $\cdots$ \\
PIFS        & Palomar & 5.0 & 51.8 & 0.45 & 115 & 0.23 & 550 & 128 & 0.22 \\
            &         & 5.0 & 51.8 & 0.45 & 115 & 0.10 & 1300 & 128 & 0.22 \\
NIFS$^a$    & Gemini  & 8.0 & 9.0 & 0.01 & 900 & 0.19 & 5300. & 1007 & $\cdots$\\
GNIRS$^a$   & Gemini  & 8.0 & 15.4 & 0.023 & 684 & 0.301 & 1700 & 512 & $\cdots$\\
            &         & 8.0 & 15.4 & 0.023 & 684 & 0.087 & 5900 & 512 & $\cdots$\\
SPIFFI      & VLT     & 8.0 & 0.54 & 0.006 & 1024 & 0.34 & 3000 & 1024 & 0.3\\
            &         & 8.0 & 10.2 & 0.001 & 1024 & 0.34 & 3000 & 1024 & 0.3\\
            &         & 8.0 & 64.0 & 0.06 & 1024 & 0.34 & 3000 & 1024 & 0.3\\ \hline
\multicolumn{10}{c}{Future Near-Infrared Instruments} \\ \hline 
KMOS$^a$    & VLT     & 8.0 & 188.0 & 0.04 & 4204 & 0.28 & 3600. & 1000 & $\cdots$\\
FISICA$^a$ & GTC & 10.4 & 72.0 & 0.53 & 136 & 0.79 & 1300. & 1024 & $\cdots$\\ 
\multicolumn{10}{c}{} \\ [-0.08in] \hline \hline
\multicolumn{10}{l}{$^a$ Advanced Image Slicer design.} \\
\end{tabular}
\end{table}

We summarize the existing and planned instruments in Table 3. The
length of the list, particularly in the planned instruments marks a
sea-change over the last few years away from fiber+lenslet coupling.
While slicers originated for NIR instruments, starting with the
now-defunct MPE-3D (Thatte et al. 1994), the list of planned optical
slicers is extensive.  Existing NIR instruments include PIFS (Murphy
et al. 1999) and UIST (Ramsay Howat et al. 2006)\footnote{See
www.jach.hawaii.edu/UKIRT/instruments/uist/uist.html for
sensitivities.} on 4m-class telescopes; NIFS (McGregor et al. 2003),
GNIRS (Allington-Smith et al. 2004), and SPIFFI (Eisenhauer et
al. 2003, Iserlohe et al. 2004), on 8m-class telescopes.  SINFONI (SPIFFI +
MACAO) on VLT (Bonnet et al. 2004) in particular has shown the power of
NIR adaptive-optics (AO) coupled to an image slicer at moderate
spectral resolution achieving 20-30\% throughput.  Future NIR
instruments include KMOS (Sharples et al. 2004) -- a multi-object
system discussed below, and FISICA. Below we also discuss three planned NIR
instruments for space.

While the only existing optical instrument is ESI (Sheinis et
al. 2002, 2006), future optical instruments include WiFeS (Dopita et
al. 2004), SWIFT (Goodsall et al., this workshop), and MUSE (Bacon et
al. 2004 and references therein). ESI is unique in being the only
cross-dispersed IFU system. While the number of spatial elements is
modest, ESI has enormous spectral multiplex (at medium spectral
resolution and good efficiency) -- the largest of any instrument
planned or in existence.

\subsection{Direct Lenslet Coupling}

This is the most significant departure in grating-dispersed 3D
spectroscopy, and therefore the most interesting. The basic concept
consists of pupil-imaging spectroscopy using lenslets. The same type
of lenslet array used in the fiber+lenslet mode create a pupil image
from each lenslet, which again is smaller than the size of the
lenslet. Here, the array of pupil-images forms the spectrograph input
focal surface, or object; no fibers or slicers reformat the telescope
focal plane into long-slit; the two-dimensional array of pupil-images
is preserved. However, the pupil image does not preserve the spatial
information within the lenslet field. These pupil images are
dispersed, and then re-imaged at the output spectrograph image
surface.

Because direct lenslet injection preserves the 2D spatial data format,
this type of instrument typically offers more spatial
coverage or sampling at the expense of spectral information.  The
extent of the spectrum from each pupil image must be truncated to
prevent overlap between pupil images. From the instrument
design perspective, what is gained is significant:
The spectrograph field of view grows linearly with $\Omega$, instead
of as $\Omega^2$ as it must in a long-slit spectrograph, where the 2D
spatial information must be reformatted into a 1D slit. Hence
this mode is best suited to instruments with the largest $\Omega$ or
$N_\Omega$.

Lenslet-coupled instruments have excellent spatial fill factor,
identical to fiber+lenslet systems, and comparable to slicers.
Because this is achieved with fewer optical elements and no fibers,
there is no information loss via FRD, and overall the system
efficiency can be very high.  As with fiber+lenslet coupling, there
are concerns about scattered light from lenslets apply here
too. Unlike fiber-coupled modes, there is no control over spatial
re-formatting. The spectra can be well-packed onto the detector, but
as noted above, the band-pass must be crafted to prevent overlap for a
given spectral dispersion, i.e., there is limited spectral coverage at
a given resolution. Spectral extraction is critical to minimize
crosstalk while maximizing $S/N$.

Existing optical systems (SAURON, Bacon et al. 2001; OASIS, McDermid
et al. 2004) have relatively low dispersion due to grism limitations,
although the grisms allow for very compact, undeviated
systems. Grating-dispersed systems do exists in the NIR (OSIRIS,
Larkin et al 2003). Future systems with VPH grisms and gratings will
have even higher efficiency; the coupling mode is well suited to
articulated-camera spectrographs. The systems summarized in Table 4
are designed to exploit superb image quality with fine spatial
sampling (OASIS and OSIRIS are coupled to AO). While they cannot take
advantage of high dispersion without becoming read-noise limited,
systems with larger specific-grasp could be optimized for high
spectral resolution.

\begin{table}
\caption{Lenslet-Coupled Integral Field Instruments}
\tiny
\begin{tabular}{llllllllll}\hline \hline 
Instrument &  Tel. & D$_T$ & $\Omega$ & d$\Omega$ &
N$_\Omega$ & $\Delta\lambda/\lambda$ & R & N$_R$ & $\epsilon$ \\
& & (m) & \multicolumn{2}{c}{(arcsec$^2$)} &  &  &  &  &  \\ \hline
\multicolumn{10}{c}{Existing Optical Instruments} \\ \hline 
SAURON & WHT  & 4.2 & 1353 & 0.88 & 1577 & 0.11 & 1213 & 128 & 0.147 \\
       &      & 4.2 & 99 & 0.07 & 1577 & 0.10 & 1475 & 150 & 0.147 \\
OASIS  & WHT  & 4.2 & 1.92 & 0.002 & 1100 & 0.50 & 1000 & 400 & $\cdots$\\
       &      & 4.2 & 31.0 & 0.026 & 1100 & 0.50 & 1000 & 400 & $\cdots$\\
       &      & 4.2 & 180. & 0.17 & 1100 & 0.50 & 1000 & 400 & $\cdots$\\ \hline 
\multicolumn{10}{c}{Future Optical Instruments} \\ \hline 
\multicolumn{10}{c}{Existing Near-Infrared Instruments} \\ \hline 
OSIRIS & Keck & 10.4 & 1.2 & 0.02 & 3000 & 0.12 & 3400 & 400 & $\cdots$\\
       &      & 10.4 & 30. & 0.10 & 3000 & 0.12 & 3400 & 400 & $\cdots$\\
       &      & 10.4 & 0.3 & 0.02 & 1019 & 0.47 & 3400 & 1600 & $\cdots$\\
       &      & 10.4 & 7.5 & 0.10 & 1019 & 0.47 & 3400 & 1600 & $\cdots$\\ \hline 
\multicolumn{10}{c}{Future Near-Infrared Instruments} \\ \hline 
\multicolumn{10}{c}{} \\ [-0.08in] \hline \hline
\end{tabular}
\end{table}

\subsection{Filtered Multi-Slit (FMS) Coupling}

The notion of direct lenslet-coupling motivates a poor-person's
alternative, which returns the riches of preserving spatial
information. The concept is to use a conventional, multi-object
imaging spectrograph with a narrow-band filter, and a slit-mask of
multi-slits in a grid pattern with grid-spacing tailored to the
desired dispersion of the grating.  This is illustrated in Figure 1.11.
Spatial multiplexing is increased via filtering. While this only
offers sparse spatial sampling, it preserves spatial information
(unlike any other mode except slicing), and can easily be adapted to
existing spectrographs.

The notion of filtering to increase spatial multiplex has been used
for multi-object spectroscopy, e.g., Yee et al. (1996) in the context
of redshift surveys using MOS on CFHT (Le Fevre et
al. 1994). Likewise, fiber+lenslet coupled IFUs, such as VIMOS and
GMOS, use filtering as an option to prevent spectral overlap in
configurations with multiple, parallel pseudo-slits; this is designed
to permit trade-offs in spatial versus spectral coverage. What is
described here is more like the multi-object mode, but instead uses a
regular grid of slitlets. This is well-suited, for example, to observing
single, extended sources.

\begin{figure}
\centering
\vspace{7.25cm}
\includegraphics{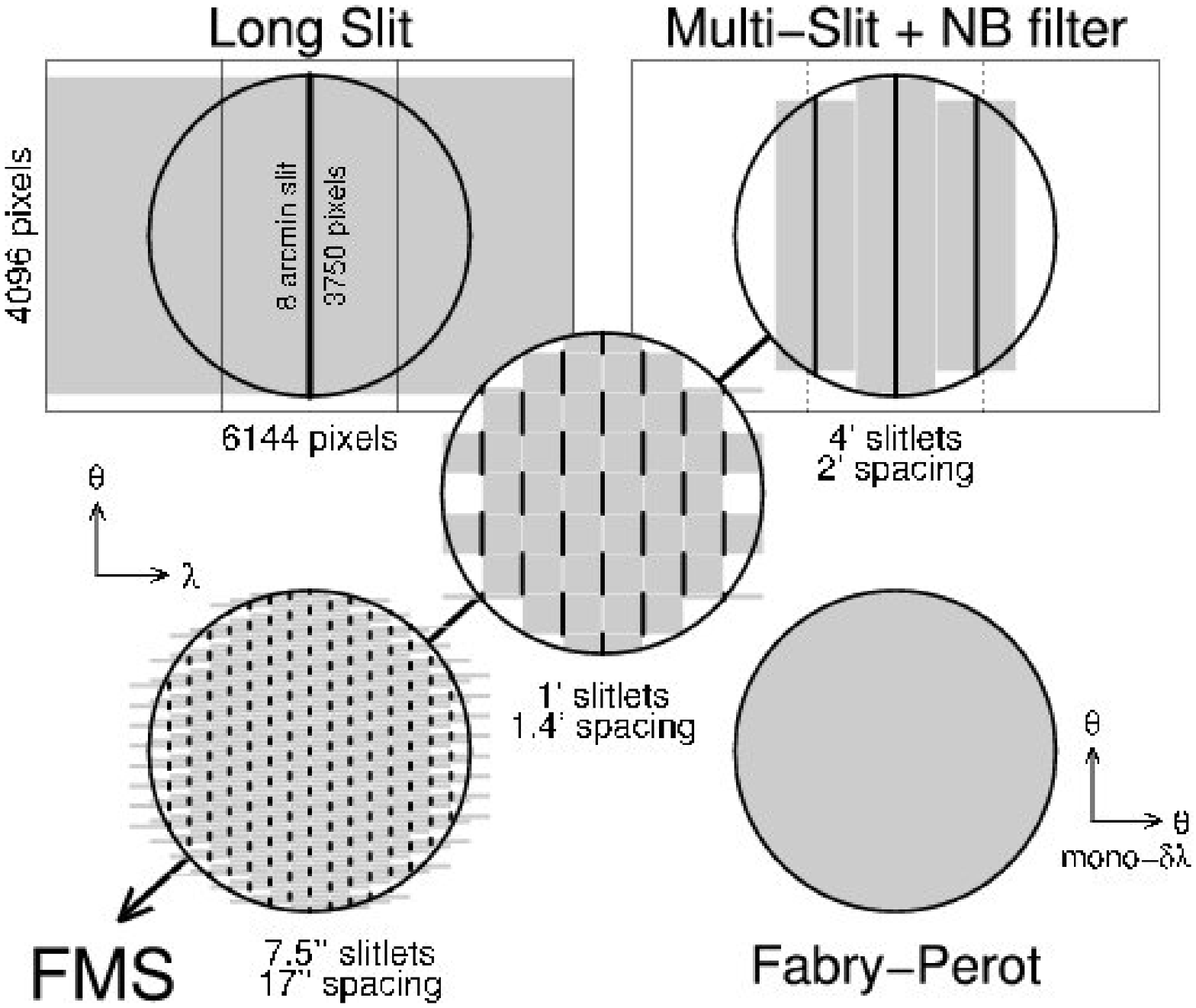}
\includegraphics{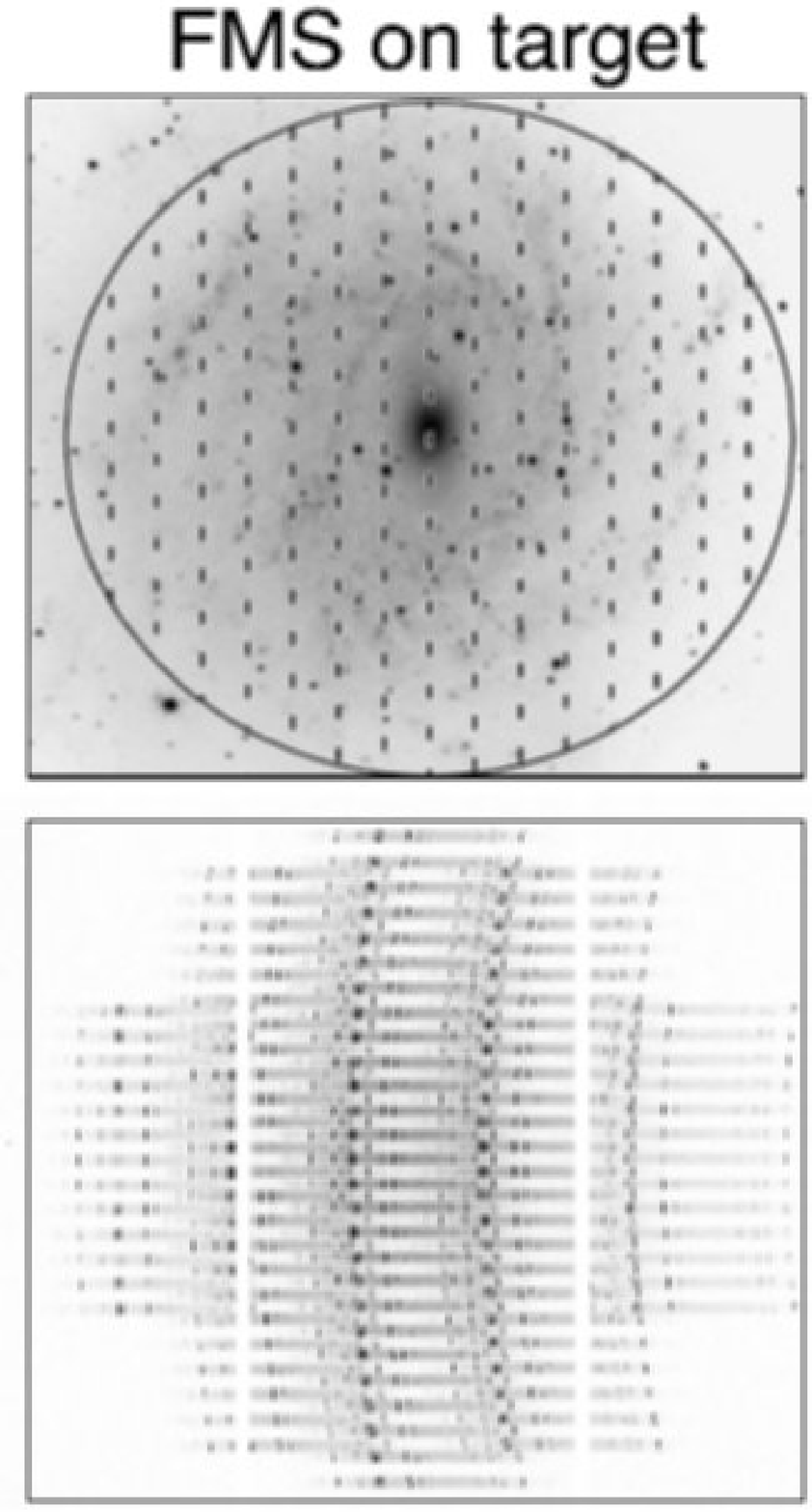}
\caption{Filtered multi-slit schematic for SALT's RSS. The 5 panels at
left show the progression from long-slit, to filtered long-slit, to
two different grids of filtered multi-slits. Both are tuned to a high
dispersion, 10 nm band-pass, and achieve a spatial multiplex gain of 3
over a pure long-slit, with a 6x loss in band-pass. Higher spatial
multiplex (2-10$\times$) is achieved at lower spectral resolution.
The RSS Fabry-Perot mode is shown for reference.  The two right-most
panels show an overlay on a nearby, face-on galaxy, and some
on-telescope calibration data for that slit-mask.}
\end{figure}

An example of this type of instrument is the SALT Robert Stobie
Spectrograph (RSS), a prime focus imaging spectrograph with an 8
arcmin field of view, articulating camera, VPH grating suite, dual
Fabry-Perot etalons, and $R = 50$ order-blocking filters (Kobulnicky
et al. 2003, Burgh et al. 2003).  The latter can be used with the
multi-slit masks to gain a factor of 3 in spatial multiplex at the
highest spectral resolutions ($R = 10,000$ with a 10 nm band-pass). At
lower resolutions (and fixed band-pass), the slit-packing can be
increased by factors of 2 to 10, such that the gain in spatial
multiplex is comparable to the loss of a factor of 5-6 in
spectral multiplex in this particular case (the system is designed for
large spectral multiplex). Even at high spectral resolution what is
gained -- beyond the spatial multiplex -- is the ability to gain 2D
spatial mapping in a single exposure. On balance, what is lost and
gained is comparable from a purely information stand-point, and hence
the choice is, as always, science-driven. For the study of nearby
galaxy kinematics, this is an outstanding approach.

\subsection{Multi-object Configurations}

Multi-object 3D spectroscopy is a major path for future
instrumentation, although it already exists today in one fabulous
instrument: FLAMES/GIRAFFE. Here we are talking about instruments with
multiple, independently positionable IFUs.  Returning to our so-called
``grand'' merit function, it is for just these types of instruments
that $\Omega_s$ is relevant.

The most obvious way to feed such an instrument is with fiber or
fiber+lenslet bundles (e.g., FLAMES/GIRAFFE). Fiber-based systems
provide flexibility for spatial positioning, but for cryogenic NIR
instruments, lenslets or slicers may be required.  This necessitates relay
optics, which are more mechanically challenging to design and build,
and introduce additional surfaces which lead to lowered throughput.
Sharples et al. (2004) have considered the multiple, deployable slicer
design for KMOS. It is also possible to implement direct lenslet
coupling (pupil imaging), as demonstrated by the MUSE concept (Henault
et al. 2004), albeit in the context of splitting up a monolithic
field into chunks fed to separate spectrographs.

\subsection{Summary of Considerations}

The various coupling methods discussed above present different
opportunities for down-selecting information, and packing three into
two dimensions in ways which trade quality versus quantity.

\subsubsection{Information Selection and Reformatting} 

Fiber+lenslet, slicers, and lenslet modes yield comparable spatial
telescope focal-surface sampling, while pure fiber systems have at
best 65\% integral coverage. Fiber-based systems, however, offer the
most flexibility in re-formatting telescope to spectrograph focal
surfaces. Slicers and FMS preserve full spatial information, but only
slicers preserve full, integral spatial information. As a result of
this coherency, slicers can give the most efficient packing on the
detector. In terms of spectral information, lenslets and FMS have
limited sampling, but other coupling modes all essentially feed
long-slit spectrographs, and therefore are comparable.

\subsubsection{Coverage versus Purity} 

Scattered light and cross-talk limit signal purity, but to avoid their
deleterious effects requires less efficient use of the detector by
e.g., broader spacing of fibers in the pseudo-slit, or band-limiting
filters, thereby limiting coverage in either or both spatial or
spectral dimensions. The trade-off optimization should be
science-driven.  Within this context, pure fiber systems and FMS
minimize scattered light, although fiber azimuthal-scrambling broadens
potential cross-talk between spatial channels of the spectrograph.
Slicer systems, again by virtue of the spatial coherency of each
slice, are able to utilize detector real-estate while maintaining
signal purity.

\subsubsection{Sky Subtraction}

There are four primary issues concerning, and root causes of,
sky-subtraction problems in spectroscopy: (i) Low dispersion:
sky-lines contribute overwhelming shot-noise.  (ii) Aberrations and
non-locality: sky-line profiles vary with field angle (spectral and
spatial) and time.  (iii) Stability: instrument-flexure and detector
fringing.  (iv) Under-sampling: compounds problems of field-dependent
aberrations and flexure. All of these conditions are further
compounded if there is fringing on CCD.

The solutions to these problems are both instrumental, observational,
and algorithmic, i.e., in the approach to the data analysis.  The
instrumental solution involves having a well-sampled, high-resolution,
and stable system (you get what you pay for).  Fiber-based systems
offer the most mapping flexibility, which is critical for
spectrographs with aberration-limited spectral image-quality. Pupil
imaging (lenslets with or without fibers) may offer advantages for
HET/SALT style telescopes, again if sky-subtraction is spectrograph
aberration-limited.

The observational approach includes (a) beam-switching, where object
and sky exposures are interleaved; and (b) nod-and-shuffle, where
charge is shuffled on the detector in concert with telescope nods
between object and sky positions. Both of these approaches have a 50\%
efficiency in either on-source exposure or in the number of sources
that can be observed (the on-detector source packing fraction).

An algorithmic approach entails aberration modeling, which is
well-suited to any of the coupling methods that feed a spectrograph in
a pseudo long-slit. The question is to what extent data analysis can
compensate for instrumental limitations and avoid inefficient
observational protocol.

Some examples exist of telescope-time-efficient sky-subtraction
algorithms -- solutions which do not require beam-switching or
nod-and-shuffle.  For example, Lissandrini et al. (1994) identify
flux- and wavelength-calibration, as well as scattered light as the
dominant problems in their fiber-fed spectroscopic data.  They use
sky-lines for 2nd-order flux calibration (after flat-fields), model
scattered light from neighboring fibers, and map image distortions in
pixel space to obtain accurate wavelength calibration. The improvement
is dramatic. Bershady et al. (2005) show that higher-order aberrations
are important; wavelength calibration is critical, but so too is line
shape.  They describe a recipe for subtracting continuum and fitting
each spectral channel with a low-order polynomial in the spatial
dimension of the data cube. The algorithm works spectacularly well for
sources with narrow line-emission with significant spectral-channel
offsets (e.g., high internal dispersion as in a rapidly rotating
galaxy, or intrinsically large velocity range, as in a redshift
survey) and well-sampled data.  For other instruments or sources (poor
sampling, low dispersion, broad lines, small velocity range): If
aberrations are significant, more dedicated sky fibers are needed. On
balance, the optical stability of the instrument is critical.

Are these post-facto, algorithmic solutions 100\% efficient? Not
quite. One still needs to sample sky, but, as derived in Bershady et
al. (2004), the fraction of spatial elements devoted to sky is
relatively low (under 10\%, and falling below 3\% when the number of
spatial resolution elements exceeds 1000). So here is a case where,
with a stable spectrograph, considerable efficiency may be gained by
employing the right processing algorithm. Consequently, fiber-fed,
bench-mounted spectrographs offer the greatest opportunities to
realize these gains. Regardless of spectrograph type and feed,
attention to modeling optical aberrations is critical for good
sky-subtraction (Viton \& Milliard 2003; Kelson 2003).

\section{Interferometry-I: Fabry-Perot Interferometry}

Fabry-Perot interferometry (FPI) provides a powerful tool for 3D
spectroscopy because FPI is field-widening relative to
grating-dispersed systems.  That is to say, higher spectral resolution
can be achieved with FPI for a given instrument beam size and entrance
aperture. This has long been recognized in astronomy. Unfortunately,
the breadth of applications of FPI to sample the data cube has been
under-utilized in astronomy. Astronomical applications almost
exclusively use F-Ps as monochromators, i.e., field-dependent, tunable
filters. This allows for a premium on spatial multiplex at the loss of
all spectral multiplex at a given spatial field-angle. Multi-order
spectral multiplex {\it can} be regained via additional grating
dispersion, as noted below -- but in astronomical applications, this
is largely a concept (with one exception). However, it is also
possible to use F-Ps for spectroscopy. In this mode, FPI yields the
converse trade in spatial versus spectral multiplex. There is again
only one example of such an existing instrument. In this sense, FPI to
date has offered two (orthogonal) extremes in sampling the data
cube. The third dimension (band-pass or field-sampling on the sky) has
been gained via the temporal domain, i.e., multiple observations. In
this sense FPI has not yet been implemented for truly 3D spectroscopy.

The basic principles of FPI, in the context of astronomical
monochromators, can be found in Geake (1959), Vaughan (1967), and many
other references.  We summarize the salient aspects to highlight here
the field-widened capabilities (we are indebted to R. Reynolds for the
structure of the formal development). We discuss and give examples of
the two different FPI applications noted above, and sketch how one
might balance spatial and spectral multiplex in future 3D instruments.

\begin{figure}
\centering
\vspace{5cm}
\includegraphics{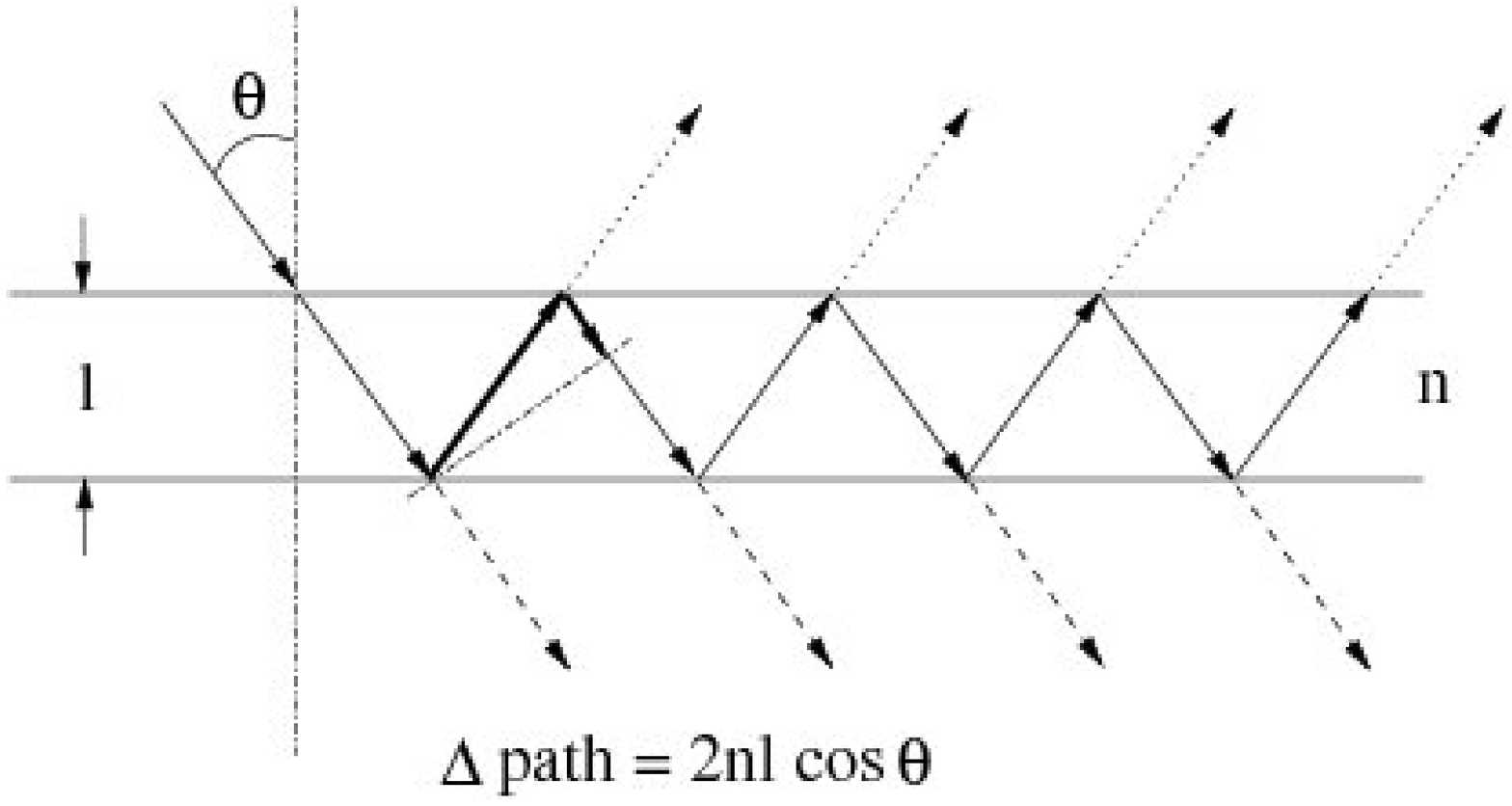}
\includegraphics{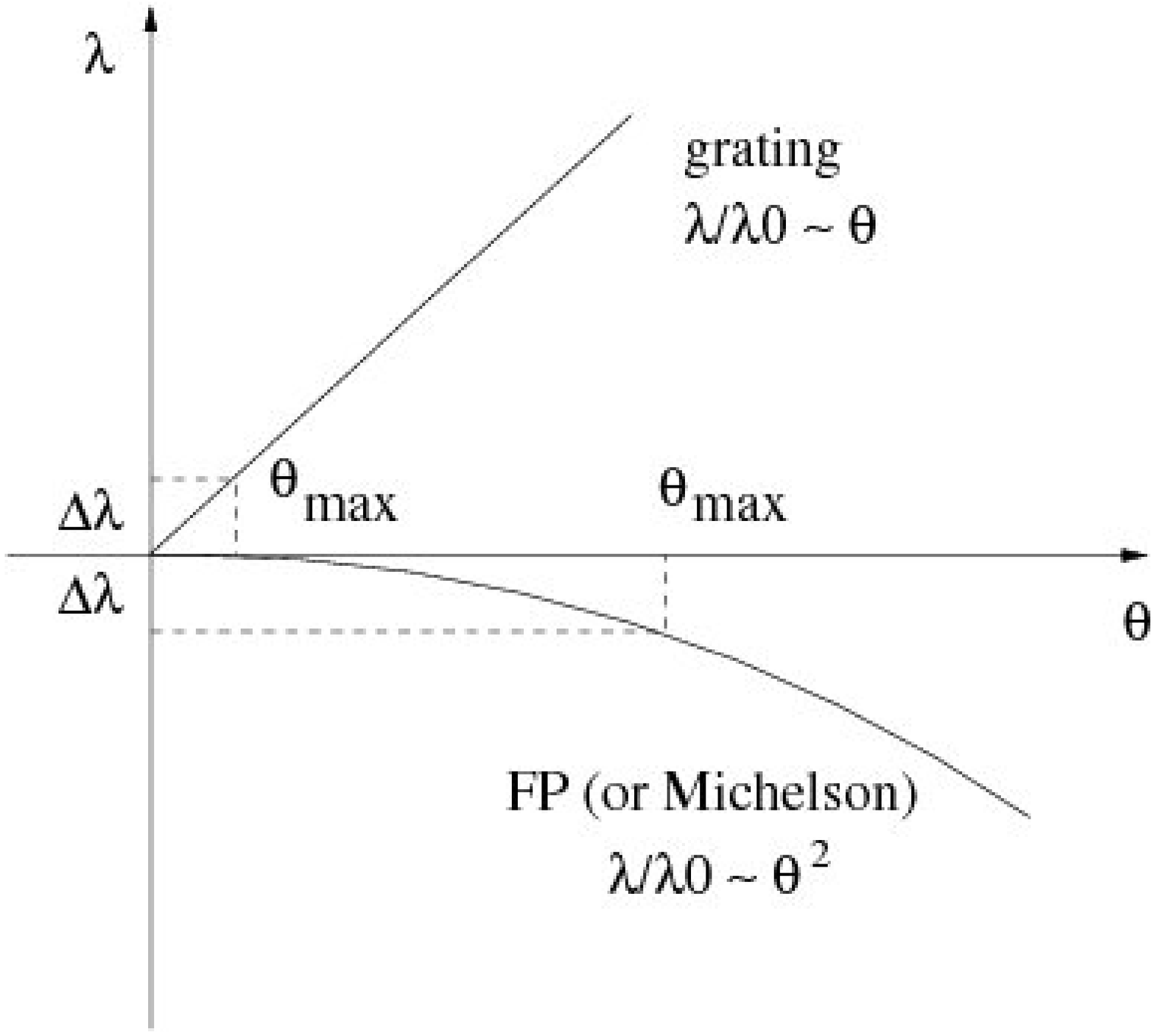}
\caption{Basic concept of etalon-interference (left), and F-P versus
grating spectral resolution as a function of angular aperture,
$\theta$ (right).}
\end{figure}

\subsection{Basic concepts and Field-Widening} 

Etalons (high-precision, flat glass plates) are parallel-spaced by some
distance $l$, filled with gas of refractive index $n$, and coated to
have high reflectivity. Light incident at some angle, $\theta$,
produces internal reflections, with transmission when the added path
($\Delta {\rm path} = 2 \ n \ l \ cos \ \theta $) between reflections
yields positive interference (left panel, Figure 1.12).  The ratio of
transmitted to incident intensity, $I_t/I_i$, is given approximately
by an Airy function with peaks ($I_t=I_i$) when $\Delta {\rm path} = m
\lambda$, where $m$ is the order. Given the geometry, this yields an
angular dependence to the transmitted wavelength: $\lambda = (2 \ n \
l \ / \ m) \ cos \ \theta$. This can be compared to the grating
equation (Littrow configurations for simplicity), where $\lambda = 2 \
n \ \Lambda_g \ / \ m \ sin \ \theta$. At small angles, this means
that the instrument entrance aperture can be larger in angle for a F-P
compared to a grating spectrograph for the same $\delta\lambda$, as
illustrated in the right panel of Figure 1.12.  In other words, a F-P
system is field-widened for the same spectral resolution (see also Roesler
1974 and Thorne 1988).

The central wavelength of the F-P is controlled via tuning the gap
($l$) or pressure (index $n$).  The free spectral range is given by
the spacing between Airy-function peaks in wavelength: $Q = 1 \ / \ 2
\ n \ l \ cos \ \theta$. Order-blocking filters are needed to suppress
other orders. Double etalons suppress the Lorentian wings in the
Airy-function.  The resolution, which is the full-width at
half-maximum of the Airy formula peak, is given by: $R = \lambda /
\delta\lambda = 2 \ n \ l \ \cos\theta \ N_\Re / \lambda = m \ N_\Re$,
where $N_\Re$ is the reflective finesse defined as $N_\Re = \pi
\Re^{1/2} / (1 - \Re)$, and $\Re$ is the reflectivity. The finesse is
equivalent to roughly the number of back and forth reflections, and
gives the number of resolution elements within the free spectral range
of the system; a typical value is $\sim$30 (see Tanaka et al. 1985 for
a more detailed discussion).  This implies that the spectral
resolution, $R$, is roughly the total path difference divided by the
wavelength. High spectral resolution requires high finesse or high
order, with the gap size tuned for the desired wavelength.  This also
achieves high contrast between the maximum and minimum transmittance
between orders: $I_{max} / I_{min} = (1+\Re)^2/(1-\Re)^2 =
1+\frac{4N_\Re^2}{\pi^2}$. Herbst \& Beckwidth (1988) provide a nice
illustration of these quantities.

\subsection{F-P Monochromators}

F-P's are conventionally thought of as being used with collimated
beams (Bland \& Tully 1989 present a review a mini-review of such
instruments from that era). In this case, there is the classic radial
wavelength dependence in the image plane. At low spectral resolution
the band-pass can be made nearly constant over a large field of view
(Jones et al. 2002), as follows.

One way to characterize an etalon is by the size of its ``bull's
eye,'' or Jacquinot spot (Jacquinot 1954). The bull's eye refers to
the physical angle $\theta$ such that $\lambda_0 /
|\lambda_0-\lambda_\theta| < R$, and is given by $\theta_{max} =
\cos^{-1}(1-1/R) \sim \sqrt{2/R}$. This quantity is independent of
the telescope, and is a property of the etalon. By coupling to a
telescope, it is possible to modify the angular scale ($\alpha$)
sampled on the sky by the bull's eye. Since $A\times\Omega$ is
conserved, $\alpha = \theta D_e / D_T$, where $D_e$ is the etalon
diameter and $D_T$ the telescope diameter.

F-P's can, however, be used in converging (or diverging) beams, even
near a focus (Bland-Hawthorn et al. 2001). Some examples include the
optical F-P on the CFHT 3.6m, when used with the AO Bonnette
(AOB)\footnote{See
www.cfht.hawaii.edu/Instruments/Spectroscopy/Fabry-Perot/, and Joncas
\& Roy (1984) for an earlier incarnation on this telescope.} and the future
F2T2, an near-infrared double-etalon system for FLAMINGOS-2 (Gemini
8m; Scott et al. 2006, Eikenberry et al. 2004a). Image information is
preserved by sampling the beam at a down-stream focus, but the
spectral resolution is lowered (for a given finesse) at any spatial
location because each field angle on the sky is mapped into a range of
physical angles through the etalon.  The degradation is not
particularly severe for lower-finesse etalons or very slow beams.  The
FLAMINGOS-2 multi-conjugate adapative optics (MCAO) focus for F2T2 is
f/30, and the AOB F-P beam is f/40. If the total angular field of view
is much smaller than the beam angle, or the focus is made telecentric,
the band-pass is constant across field angles on the sky, and the
system forms a highly uniform tunable filter.  The AOB optics are not
telecentric; this produces a radial degradation in the resolution.

\subsection{F-P Spectrometers}

Alternatively, the full spectral information can be extracted at the
loss of the spatial information by placing the etalons at or near a
telecentric focus and sampling the pupil in a collimated beam. The
Wisconsin H$\alpha$ Mapper (WHAM; Reynolds et al. 1998) is the only
astronomical example of this type of instrument.  In this instance,
the light is collimated {\it after} it passes through the etalons,
never refocused, and a detector is placed at the pupil formed by the
collimator. Field position on the detector contains spectral
information: each radius corresponds to a different wavelength. This
is similar to the monochromator application, except in this case each
radial location on the detector has a superposition from all
spatial locations on the sky within the instrument entrance aperture.

\subsection{3D F-P Spectrophotometers}

\subsubsection{Grating-Dispersed FPI} 

Arguably the most interesting F-P monochromator mode is to eliminate
the order-blocking filters, and grating-disperse the output beam to
separate the orders onto the detector to increase the spectral
multiplex. See, for example, le Coarer et al.'s (1995) description of
PYTHEAS. Baldry et al. (2000) work out a particularly compelling case
for a cross-dispersed echelle system.  The gain in spectral multiplex
does not necessarily cost spatial multiplex. In practice, some F-P's
are in spectrographs where they under-fill the detector and usable
field in the image plane (e.g., RSS and F2T2). If the dispersion is
significantly greater than the etalon resolution, then in addition to
spectral multiplex, this mode adds band-limited slitless spectroscopy
in each F-P order.

\subsubsection{Pupil-Imaging FPI} 

The above discussion frames the notion that detection down-stream of
an etalon at the pupil of a collimated beam provides spectral
information but no spatial information, while detection at a focal
surface provides the complement. A simple ray-trace shows that between
these two locations spectral and spatial information are mixed.  By
using pupil imaging at the system input via a lenslet array (\S
1.2.9), detection at an intermediate surface in a converging beam can
separate spatial and spectral information. Although this has never
been done, in principle this could balance spatial and spectral
multiplex and allow for true 3D spectroscopy in future, field-widened
instruments.

\subsection{Sky Stability} 

Because spectral channels are not observed simultaneously in
monochromatic modes, atmospheric changes must be calibrated (see, for
example, Atherton et al. 1982 in the context of TAURUS). Field stars
may suffice if they are sufficiently featureless over the scanned
wavelength range. Built-in calibration is desirable, which can be
achieved, for example, via a dichroic feeding a monitoring camera.
This capability is designed for new generation of instruments (e.g.,
ARIES, T. Williams, private communication).

\subsection{Examples of Instruments}

Two extremes in F-P instrumentation are highlighted by the RSS imaging
F-P (Williams et al. 2002) and the WHAM non-imaging F-P.  Both have
150 mm etalons, but the RSS system is coupled to a 9.2 m telescope
with an 8 arcmin field of view, 0.2 arcsec sampling and spectral
resolutions of 500, 1250, 5000, and 12,500. In contrast, WHAM is
coupled to a 0.6m telescope, with a 1 deg field of view {\it and} angular
resolution, spectral resolution of $R = 25000$, and spectral coverage
of about 166 resolution elements for one spatial element.

There are a large number of existing F-P monochromators (a.k.a.,
tunable filters), indicated even by the following incomplete list.
Optical systems include, but are not limited to: PUMA (OAN-SPM 2.1m,
Rosado et al. 1995), RFP (CTIO 1m and 4m; e.g., Sluit \& Williams
2006), CIGALE (ESO 3.6m and OHP 1.9m; Boulesteix et al. 1984), FaNTOmM
(OMM 1.6m, OHP 1.9m, and CFHT 3.6m; Hernandez et al. 2003), Goddard F-P
(APO 3.5m; Gelderman et al. 1995), SCORPIO F-P (SAO 6m, Afanasiev \&
Moiseev 2005), IMACS F-P (Magellan 6.5m; Dressler et al. 2006), as well
as the above-mentioned CFHT F-P etalons which can be used with the AOB
as well as the MOS and SIS systems.  The most widely cited system is
TTF/TAURUS-II (AAT 3.9m, WHT 4.2m; Gordon et al. 2000 and references
therein).  Existing infrared instruments include NIC-FPS (Arc 3.5m;
Hearty et al. 2004), GriF (CFHT 3.6m; Clenet et al. 2002), PUMILA
(OAN-SPM 2.1m, Rosado et al. 1998), UFTI (UKIRT 3.8m, Roche et
al. 2003) and NACO (VLT 8m; Hartung et al. 2004, Iserlohe et
al. 2004). GriF, NACO, and F2T2 are AO-fed. By virtue of their use in
collimated beams, many of the F-P systems are designed to be
transportable between instruments (i.e., spectrographs or
focal-reducers) and telescopes.  Future instruments include the
optical OSIRIS (GTC 10.4m) and near-infrared FGS-TF (JWST 6.5m; Davila
et al. 2004) and F2T2 (above).  These systems span a wide range of
wavelength, spectral, and spatial resolution. One attribute they have
in common is a spectral multiplex of unity.

\section{Interferometry-II: Spatial-Heterodyne Spectroscopy }

A spatial-heterodyne spectrometer (SHS) is a Michelson interferometer
with gratings replacing the mirrors. The principles of operation are
described and illustrated by Harlander et al. (1992) -- a paper
well-worth careful study.\footnote{The presentation here benefited
  from discussion w/ J. Harlander, A. Sheinis, R. Reynolds,
  F. Roesler, and E. Merkowitz.} Briefly, each grating diffracts light
at wavelength-dependent angles. Because of the 90-degree fold between
the two beams, the wavefronts at a given wavelength are tilted with
respect to each other after beam recombination. This tilting produces
a sinusoidal interference pattern with a frequency dependent on the
tilt angle. The degree of tilt is a function of wavelength, simply due
to the grating diffraction, and hence the interference pattern
frequency records the wavelength information.

It is easiest to conceptualize this in terms of two identical gratings
(as illustrated by Harlander et al. in their Figures 2 and 3), but in
principle the gratings do not need to be the same. Wavefronts produce
interference patterns with frequencies set by wavelength, with the
central wavelength producing no interference.  Hence the signal is
heterodyned about the frequency of the central wavelength.  Resolution
is set by the grating aperture diameter because this sets the
wavelength (i.e., angular tilt) which minimally departs from the
central wavelength which can produce the first (lowest) frequency for
interference. Bandwidth is set by the length of the detector, i.e.,
how many frequencies can be sampled depends on the number of pixels.

The advantage of an SHS over a Michelson is that no stepping is
required to gain the full spectral information, but the field of view
is reduced. The SHS can be fed with a long-slit or lenslet array,
although with the latter a band-limiting filter is needed (as with a
conventional dispersed spectrograph).  Like with a Michelson, however,
field-widening is possible via prisms. In the SHS application, the
prisms give gratings the geometric appearance of being more perpendicular
to the optical axis, and hence larger field angles are mapped within
the beam deviation producing the lowest-order interference fringe.
Cross-dispersion is possible (by tilting one of the gratings about the
optical axis), but the same fundamental limits apply concerning 3D
information formatted into a 2D detector!

One of the problems with the standard Michelson or SHS interferometer
is that their geometry throws out half the light right from the start.
Non-lossy geometries are possible.  Harlander et al. (1992) give an
example of working off-axis on the collimating mirror (see their
Figure 5). This is a perfect application for holographic
gratings. Transmission-grating geometries would eliminate the need to
go off-axis and probably allow for larger field.  Another approach is
a Mach-Zender style interferometer (Douglas 1990). The latter requires
twice the detector real-estate for the same number of spectral
resolution elements.

The primary advantage of an SHS is that it allows for very high
spectral resolution for a given solid angle relative to a
conventional, grating-dispersed spectrograph. The SHS is field-widened
like a F-P. This means the SHS can be built for low cost even on large
telescopes because the optics are small.

% this is mis-leading: diffraction-limited high-resolution capability

However, because the signal is in the form of an interferogram, there
is what is known as the ``multiplex disadvantage.'' This can be
expressed as the $S/N$ performance of the SHS relative to a grating
spectrograph: $S/N_{SHS} = S/N_{GS} (f/2)^{1/2} (S_{SHS}/
S_{GS})^{1/2}$, where $S/N_{SHS}$ and $S/N_{GS}$ are the signal to
noise in SHS and grating spectrometer, respectively, $S_{SHS}$ and
$S_{GS}$ are the total photon signal, respectively, and $f$ is the
fraction of total signal in a given spectral channel ($f < 1$, and
decreases with bandwidth). In words, this means that an SHS looses
competitiveness with grating-dispersed spectrographs when the
band-pass is large.  This has implications for design and use.
Clearly one must make $S_{SHS}$ and $f$ as large as possible.  The
small, compact optics of a SHS system lend itself to efficiency
optimization.  To make $f$ as large as possible, one must choose a
small band-width (but more than a Fabry-Perot monochromator!)  and
remove OH lines via pre-filtering, or by selecting band-passes between
them.  Returning to Figure 1, SHS is between a F-P monochromator and
other IFS methods, and therefore will have application to a broad
range of science programs that seek high spectral resolution over a
limited band-pass with good spatial coverage.

\section{Summary of Existing Instruments}

\begin{figure}
\centering
\vspace{4.25cm}
\includegraphics{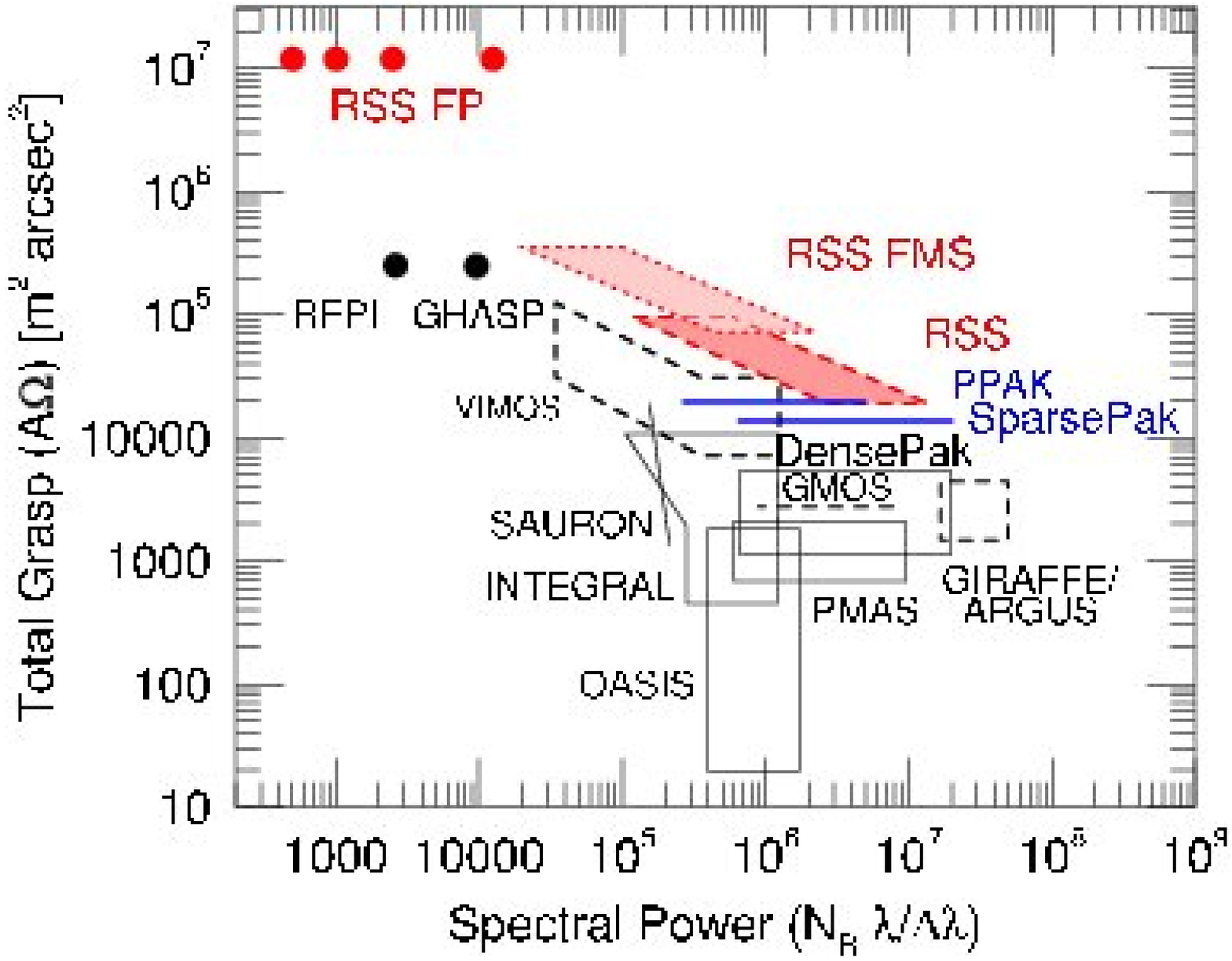}
\includegraphics{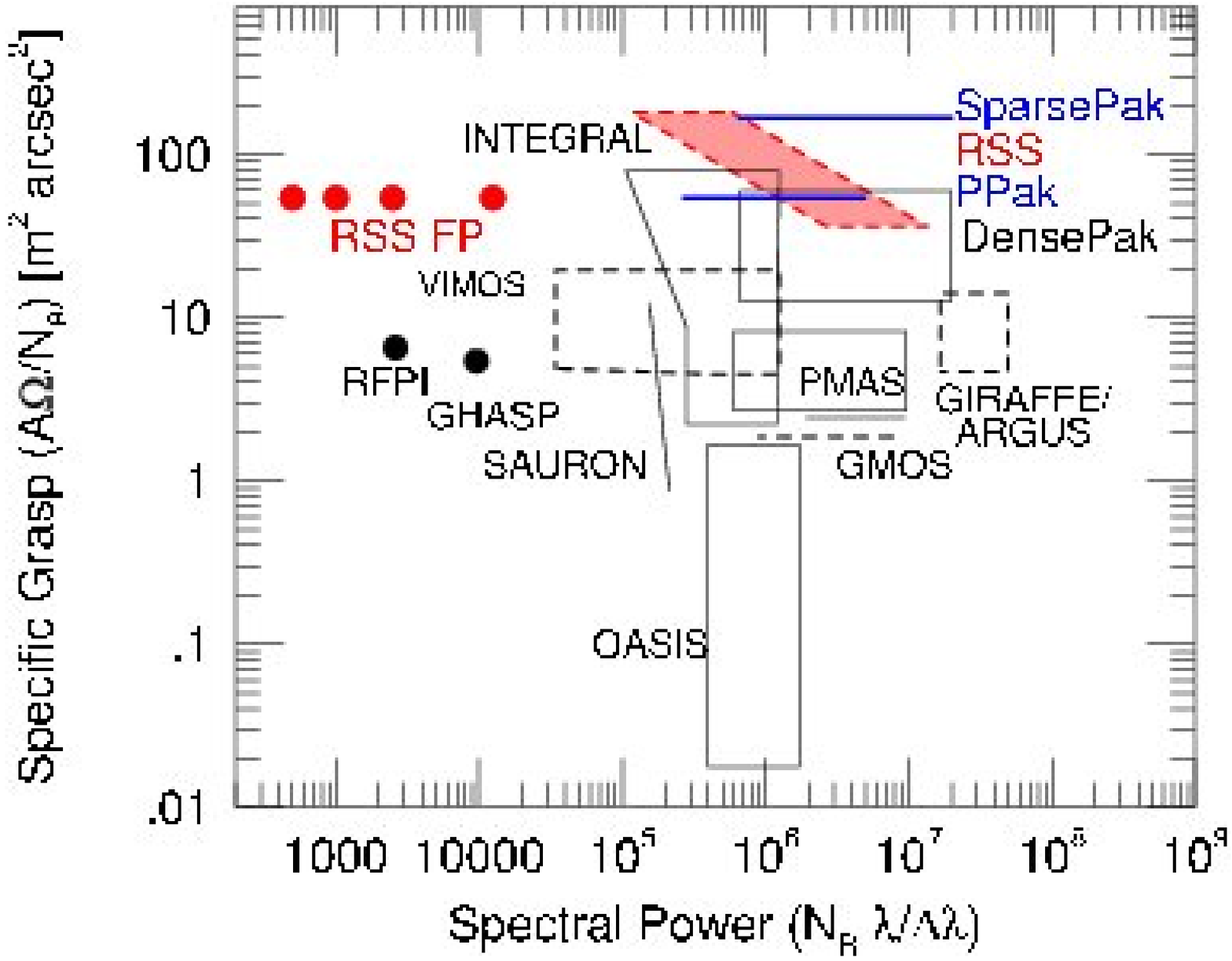}
\caption{Total and Specific Grasp versus Spectral Power for a range of
instruments on 4m- and 10m-class telescopes (solid and dashed lines,
respectively) partially updated from Bershady et al. (2005). See text
for comments on instrument efficiency.}
\end{figure}

Here we explore the sampled parameter space in spatial versus spectral
information, as well as coverage versus resolution, starting with
grasp and spectral power (Figure 1.13).  Recall that because reliable,
consistent measurements of efficiency are unavailable for most
instruments, we use grasp instead of etendue ({\tt warning:} we really
want etendue).  Note, however, that there is a factor of 6 range in
the known efficiencies of instruments tabulated in this Chapter.
Further note that there are two ways of viewing the specific
grasp. From the perspective of staying photon-limited at high spectral
resolution, high specific grasp is important. The ``flip side'' is
that low specific grasp implies high angular resolution.

Figure 1.14 shows that spatial resolution is higher in NIR
instruments, while spectral resolution is higher in optical
instruments.  Fiber IFUs have the largest specific grasp -- reflected
in the bifurcation seen in spatial resolution, i.e., fiber-fed
instruments have large footprints per element ($d\Omega$). There is a
trend of decreasing specific grasp going from fiber+lenslet, lenslet,
and finally to slicers. ESI has unusually large $A \ \times \ d
\Omega$ for a slicer; RSS in FMS mode has the highest specific grasp
overall.

\begin{figure}
\centering
\vspace{5.5cm}
\includegraphics{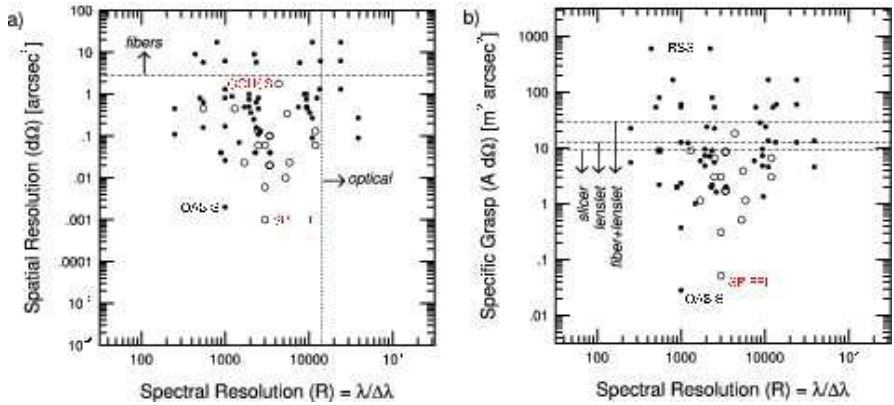}
\caption{Spatial resolution (a) and specific grasp (b) versus spectral power for
all instruments in Tables 1-4, highlighting differences between optical (filled
symbols) and NIR (open symbols), as well as
between different coupling methods (labeled).}
\end{figure}

Figure 1.14 and 1.15 together show that optical and near-infrared
instruments trade spatial resolution for grasp; there are no
high-grasp NIR instruments; the highest spectral power instruments are
optical. Optical and near-infrared instruments sample comparable total
information, with optical instruments sampling a broader range of
trades between spatial versus spectral information.  Older NIR
instruments clearly suffer from being detector-{\it size} limited.
IMACS-IFU stands out as having significantly larger number of total
information elements, $N_R \ \times \ N_\Omega$, and in this sense is
on-par with future-generation instruments.

\section{The Extended-source Domain}

One area of extra-galactic science is clearly under-sampled by
existing instrumentation, namely high spectral-resolution yet low
surface-brightness 3D spectroscopy of extended sources. The scientific
impetus is for detailed nebular studies (ionization, density,
metallicity, abundances) of not only compact HII regions, but to
extend such study to the diffuse ionized gas. Likewise, a significant
fraction of the stellar light in galaxies is in extended distributions
at low surface-brightness, i.e., below the night-sky background. The
kinematic and chemical properties of these stars is largely unknown
outside of resolved populations in the Local Group. Stellar kinematics
of galaxies on spatially-resolved scales are required to dissect the
mass distribution and detailed dynamics of disk, bulge, and halo
components. This information is effectively the Rosetta Stone for
deciphering how galaxies have assembled.

One concern with most existing IFU spectrographs is their focus on
very fine spatial sampling. Referring back to Figure 1.4, on
telescopes as small as only 10m ({\it !}), this severely limits the
spectral resolution that can be achieved at sub-arcsec sampling in the
photon-limited regime. For example, FLAMES/GIRAFFE is unusual in its
high spectral resolutions of 10-40,000.  Each IFU unit is a $2 \times
3$ arcsec of 20 rectangular microlenses sampling only $0.52 \times
0.52$ arcsec; this is equivalent to a 1 arcsec fiber on 3.5m
telescope. The instrument is very close to the
photon-detector--limited divide. The IMACS-IFU should be in a similar
domain at its high spectral-resolution limit.

\begin{figure}
\centering
\vspace{5.5cm}
\includegraphics{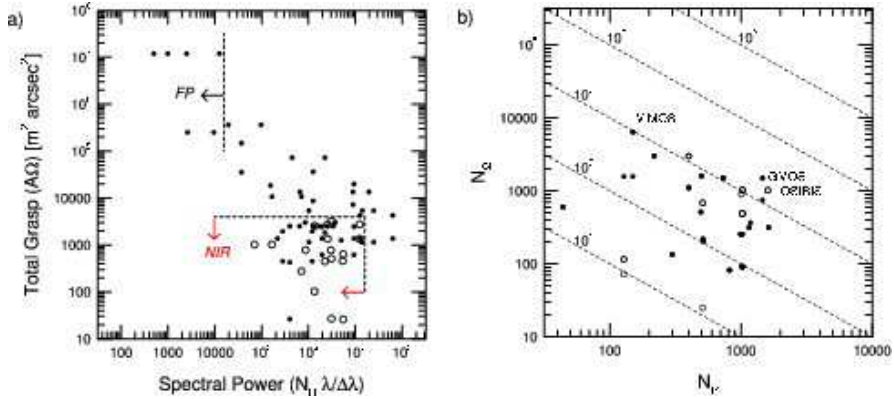}
\caption{Total grasp versus spectral power (a) and the number of
spatial ($N_\Omega$) versus spectral ($N_R$) resolution elements (b) for all
instruments in Tables 1-4,  highlighting differences between optical (filled
symbols) and NIR (open symbols). Dashed lines are at constant (labeled) total
information ($N_R \ \times \ N_\Omega$).}
\end{figure}

There is no question FLAMES/GIRAFFE has proven spectacular for
emission-line work, particular if line-emission is clumpy and
unresolved, e.g., ionized-gas kinematics of distant galaxies (Flores
et al. 2004).  The need for high angular resolution in the
distant-galaxy kinematic game {\it is} paramount. Even with $\sim$0.5
arcsec resolution, HST images are needed to super-resolve the IFU data
(Flores et al. 2004). It will be difficult, however, to use this same
facility to study diffuse gas or the stellar continuum in resolved
sources.  Furthermore, resolved structures at high redshift are all at
{\it apparently} low surface-brightness because of cosmological
dimming. To stay photon-limited, observing in the
low-surface-brightness regime requires either lower spectral
resolution, larger apertures ($d\Omega$), or larger telescopes.  Will
this be addressed by future instrumentation?

\section{Future Instruments}

The next generation of instruments will compete on both space-based
platforms such as the JWST, and on ground-based telescopes reaching
30m or larger in diameter. Why build these bigger telescopes? The
argument of simply collecting more photons is compelling but not
sufficient. New facilities, which come at increasingly greater cost,
must yield gains above the linear increase in area. Such ``windfalls''
may include over-coming detector-noise, the diffraction-limit (at long
wavelengths), backgrounds (in the case of space-based platforms), or
critical combinations thereof.

A discussion of backgrounds and the relative merits and niches of
8m-class space-based telescope such as JWST, and large 30m-class
ground-based telescopes was vetted in the early planning stages of
what was once known as ``MAXimum Aperture Telescopes,'' or MAXAT.
Gillett \& Mountain (1998)\footnote{See also the AURA MAXAT Final
Report (1999), \\
www.gemini.edu/science/maxat/maxat2\_final\_report.pdf.}  pointed out
that a cooled space-craft has significantly lower background in the
infrared compared to the ground -- even at high spectral
resolution. This contrast is dramatic for $\lambda > 2.5 \mu m$, i.e.,
in the thermal-IR. However, they calculated that above $R \sim 1000$,
8m-class space-telescopes are detector-limited at any wavelength,
assuming 0.05 arcsec apertures, a generous system throughput, and
realistic detectors. They constructed a competitiveness criteria which
assumed diffraction-limited performance for stellar imaging or
spectroscopy. Compared to JWST, they concluded ground-based telescopes
can be competitive at $\lambda < 2.5 \mu m$ for imaging if $D_T >
20m$, and for diffraction-limited spectroscopy at $R>1000$ for
$D_T>8m$ at any wavelength. These considerations have been influential
on the planning and design of future-generation instruments in the era
of JWST.

\subsection{Ground-based Instruments on 10m Telescopes}

Given rapid growth in 3D spectroscopy, we expect many new and
retro-fitted systems in the coming years.  We sketch three instruments
-- MUSE, VIRUS, and KMOS -- because they highlight the common themes
of object and instrument multiplexing. (US scientists will note two of
these systems are on the VLT.) Object multiplexing is a departure for
3D instrumentation; instrument multiplexing is a departure
overall. The basic parameters of these 3 instruments are summarized in
Tables 1,3.

Both MUSE and VIRUS offer unprecedented spatial sampling. MUSE
provides a truly integral 1 arcmin$^2$ area, sampled at the 0.2 arcsec
scale, accomplished via image slicing (AIS-type). The most significant
portions of the system are the slicers, which must perform well (with
little scattered light) in the optical, and the field-partitioning 
between a bank of 24 identical spectrographs. In comparison, the
individual spectrographs are modest, albeit high-efficiency,
articulated VPH-grating systems.

VIRUS uses the same notion of a replicated spectrograph unit (also
articulated VPH-grating systems), but in this case fed by bare fibers
at a much coarser scale ($25\times$ larger $d\Omega$).  The field
sampling is sparse, and hence this instrument follows directly in the
path of bare-fiber IFUs. What stands out in the VIR{\it U}S design is
the ``{\it U}ltra-cheap'' notion of the spectrograph unit, i.e., by
building many, replicated units, costs are lowered by economies of
scale. Such a demonstration has important implications for future
large-telescope instrument design (how large must the replication
scale be to manifest significant economy?). If the full replication of
132 spectrographs is accomplished, this will be by far the
widest-field (largest grasp) IFU in existence, likely for years to
come.  To achieve the VIRUS-132 goal requires an all new, wide-field
prime-focus spherical-aberration corrector yielding a 16 arcmin
science-grade field for the HET -- a significant opto-mechanical challenge
in itself.

MUSE, in contrast to VIRUS, has 9$\times$ less total grasp, but
almost 3$\times$ more spatial elements ($N_\Omega$). In other words,
both stand out as remarkable in spatial sampling in their own way. The
differences in spatial resolution versus coverage between MUSE and VIRUS
lies in their respective science themes. VIRUS is designed as a
precision-cosmology engine to measure the baryon oscillations by
detecting $z \sim 3$ Ly$\alpha$-emitters, and using their distribution
as a density tracer. These sources are relatively rare (in
surface-density to a given detected flux), although the exact
flux-density relation is still uncertain. Rather low spectral
resolution ($R<1000$) is needed, since only line-identification (in
the blue where backgrounds are low) and redshifts are required.

MUSE, in contrast, is designed to probe the detailed internal
properties (dynamics, stellar populations) of galaxy populations over
a wide range in redshift and in a representative cosmological
volume. The aim of this instrument is essentially to enable
spectroscopic versions of many ``Hubble Deep Fields,'' each with
sufficient spectral and spatial information to extract kinematics and
line-diagnostics of many thousands of $z<1$ galaxies.

KMOS has much the same science goals of MUSE, with the key distinction
of pushing to higher redshifts by using the NIR to capture the optical
rest-frame. By pushing to higher redshift to gain temporal leverage 
on the galaxy formation and evolution process, the source-distribution
becomes apparently fainter, and the NIR backgrounds are
higher. Consequently, on the same size telescope, one is forced to
look at intrinsically more luminous and hence rarer objects. Therefore, the
KMOS design moves away from the notion of a monolithic integral-field,
to a 24-probe system in a large, 7.5 arcmin diameter patrol field.
Each probe spans a $2.8 \times 2.8$ arcsec area sampled at $0.2 \times
0.2$ arcsec. While a multi-IFU instrument already exists (again on the
VLT) in the optical with FLAMES/GIRAFFE, the extension to the NIR
using slicers with twice the number of probes will be a significant
technical achievement.

What is missing from this suite of remarkable instruments is a design
which pushes forward a significant increase in spectral sampling
(spectral power) or specific grasp. For example, none of these
instruments offers over $R=4000$ and $N_R=2000$. This means, for
example, that advances in the study of low surface-brightness,
dynamically cold ($\sigma < 80$ km/s) systems or nebular regions will
require additional instrument innovation.

\subsection{Ground-based Instruments on 30-50m Class Telescopes}

\begin{table}
\caption{Future TMT Integral Field Instruments}
\tiny
\begin{tabular}{llllllllll}\hline \hline
Instrument & Coupling & D$_T$ & $\Omega$ & d$\Omega$ &
N$_\Omega$ & $\Delta\lambda/\lambda$ & R & N$_R$ & $\epsilon$ \\
 &  & (m) & \multicolumn{2}{c}{(arcsec$^2$)} & &  &  &  &  \\ \hline
IRMOS & slicer    & 30. & 40. & 0.01 & 4000 & 0.25 & 2000 & 500 & $\cdots$ \\
IRMOS & slicer    & 30. & 40. & 0.01 & 4000 & 0.25 & 10000 & 2500 & $\cdots$ \\
IRIS & slicer     & 30. & 0.26 & 1.6e-5 & 16384 & 0.05 & 4000 & 200 & $\cdots$ \\
IRIS & slicer     & 30. & 1.33 & 8.1e-5 & 16384 & 0.05 & 4000 & 200 & $\cdots$ \\
IRIS & slicer     & 30. & 7.93 & 4.8e-4 & 16384 & 0.05 & 4000 & 200 & $\cdots$ \\
IRIS & slicer     & 30. & 41.0 & 2.4e-4 & 16384 & 0.05 & 4000 & 200 & $\cdots$ \\
WFOS & fiber+lens & 30. & 810. & 0.56 & 1440 & 1.37 & 5000 & 6850 & 0.3 \\ 
\multicolumn{10}{c}{} \\ [-0.08in] \hline \hline
\end{tabular}
\end{table}

We summarize some of the specific exmaples of TMT 3D-spectroscopic
instrumentation in Table 5, based on D. Crampton's overview (Ringberg
2005; Crampton \& Simard 2006).\footnote{See also
www.tmt.org/tmt/instruments.}  TMT instrument design is largely
driven by AO capabilities, where the salient point is that there are
many ``flavors'' of AO, with associated levels of difficulty and risk
(inversely proportional to their performance in either image quality,
field of view, or both). The IFU-capable TMT instruments include, in
order of decreasing AO requirements: (i) IRMOS, a NIR multi-object
integral-field spectrograph fed by the multi-object adaptive object
system (MOAO), capable of 20 positional, 5 arcsec compensated patches
within a 5 arcmin patrol field; (ii) IRIS, a NIR imager and integral
field spectrograph working at the diffraction limit, fed by the
narrow-field facility AO system (NFIRAOS); and (iii) WFOS, an optical,
wide-field, seeing-limited spectrograph with potential for
a modest-grasp IFU with good spectral power and spectral resolution ($R
< 6000$).

With the exception of WFOS, instrument design is driven by AO
considerations because of the enormous physical size of the image
(which scales with mirror diameter for a constant f-ratio). WFOS is
necessarily a monster. The AO-driven focus is suitable for scientific
studies of un- or under-resolved sources (stars, planets, sub-kpc
scales in distant galaxies), and excellent science-cases have been
developed. Of this excellence there is no doubt. Of concern is that
once wedded to the notion of a very large telescope with no clear path
to building affordable, comparably-monstrous instruments, one is
forced down a path, {\it ab initio}, of considering {\it only} science
enabled by high-angular resolution.  It is not surprising to note that
WFOS -- the one non-AO corrected instrument -- stands out as also the
one TMT instrument concept that breaks into the high specific-grasp
domain at modest spectral-resolution domain (Figure 1.16). Indeed, as
seen in Figure 1.17, WFOS breaks new ground in terms of its total
grasp at the highest spectral power of any existing instrument (save
ESI). To optimize low-surface-brightness studies, other paths will need to be forged
to push to higher spectral power and resolution at comparably high
grasp.

These same trends are also being played out for instrument design for
ELT (e.g., Eisenhauer et al. 2000, Russell et al. 2004). We've focused
on TMT because of the more mature stage of this telescope's
planning. No doubt ELT's complement of instruments will open up
exciting new capabilities, as demonstrated by the superb, forefront
instrumentation on the VLT. The TMT instrumentation program, like that
of the European ELT, is evolving rapidly. What is presented here is a
snapshot circa late 2005.

\subsection{Space-based Instruments}

\begin{table}
\caption{Future Space-Based Integral Field Instruments}
\tiny
\begin{tabular}{lllllllllll}\hline \hline
Instrument & Coupling & Tel. & D$_T$ & $\Omega$ & d$\Omega$ &
N$_\Omega$ & $\Delta\lambda/\lambda$ & R & N$_R$ & $\epsilon$ \\
 & &  & (m) & \multicolumn{2}{c}{(arcsec$^2$)} & &  &  &  &  \\ \hline
FGS-TF & FP & JWST & 6.5 & 38088. & 0.018 & 2.10e7 & 0.01 & 100 & 1 & $\cdots$ \\
NIRSpec & AIS & JWST & 6.5 & 9. & 0.0056 & 1600 & 0.34 & 3000 & 1024 & $\cdots$ \\
MIRI & AIS & JWST & 6.5 & 51.8 & 0.30 & 173 & 1.48 & 2800. & 4096 & $\cdots$ \\
SNAP-IFU & AIS & SNAP & 2. & 9.0 & 0.022 & 400 & 1.95 & 100 & 195 & 0.44\\ 
\multicolumn{11}{c}{} \\ [-0.08in] \hline \hline
\end{tabular}
\end{table}

We summarize the planned 3D-spectroscopic instruments for JWST and
SNAP (Super Nova Acceleration Probe; Aldering et al. 2002) in Table
1.6. (There are other missions, which include IFUs, also in the
planning stages.)  On JWST, in remarkable constrast to HST, three of
the four instruments have 3D spectroscopic modes in the near- and
mid-infrared: (i) FGS-TF (of which F2T2 is the ground-based analogue)
delivers a $2.3 \times 2.3$ arcmin field at $R \sim 100$, with two
cameras covering 1.2 to 4.8 $\mu$m. (ii) NIRSpec (Prieto et al. 2004)
has a $3 \times 3$ arcsec IFU using and AIS with 40 $3 \times 0.075$
arcsec slices, covering 0.8-5 $\mu$m at $R=3000$.  (iii) MIRI (Wright
et al. 2004) has 4 simultaneous image-slicers at $R \sim 3000$ feeding
4 wave-bands between 5-28 $\mu$m. Each samples $4.6 \times 5.5$ arcsec
(increasing by a factor of two between bluest and reddest channel) with
an 0.37 arcsec slit-width (changing by a factor of 4 between bluest and
reddest channel).  Quoted numbers represent mean values over all
channels.

The SNAP IFU (Ealet et al. 2003) is designed to identify SNe type out
to $z\sim1.7$.  As such, it is unique in being dual optical-NIR
systems (0.35-1.7 microns), with a $3 \times 3$ field using AIS, but
very low spectral resolution ($R = 100$). With its very high expected
efficiency, coadded data-sets should yield superb, spatially resolved
spectrophotometry of galaxies on 1-2 kpc scales.

\begin{figure}
\centering
\vspace{5.5cm}
\includegraphics{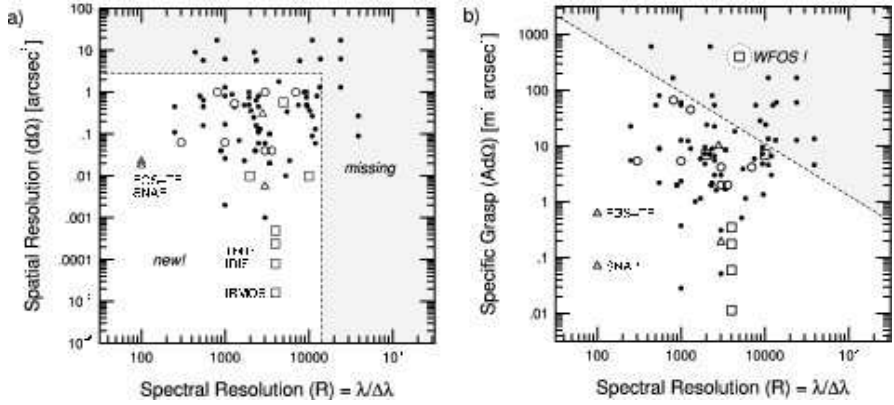}
\caption{Spatial resolution (a) and specific grasp (b) versus spectral power including
future instruments in Tables 5 and 6: existing (filled circles),
future ground-based 10m-class telescope (open circles), future
TMT (open squares), and future space-based (open triangles).}
\end{figure}

Overall, future space-based capabilities can be characterized as
having $3 \times 3$ arcsec fields mapped with AIS-technology with 0.15
arcsec sampling -- lower spatial resolution than TMT. Spectral
resolution is in the $100<R<3000$ range, again lower than TMT.  This
is consistent with their being competitive in performance relative to
TMT-class instruments, given Gillett \& Mountain's (1998) argument.
However, there are no large-grasp systems that take full advantage of
the low backgrounds of space. There are no high- or even
medium-resolution spectrographs to couple, competitively, to such
large angular apertures.  Nonetheless, barring past fiascos, the
space-based missions offer the guarantee of superlative image quality
and low backgrounds extending into the mid-IR, while ground-based
observatories face the intense challenge of developing advanced AO
systems.

\subsection{Summary of Future Instruments}

While Figure 1.16 shows some of the areas {\it not} accessed by currently
planned future instrumentation, at the same time clearly great strides
are planned for accessing new domains in spatial resolution -- {\it
from the ground.} This is encouraging because only with the largest
apertures can we stay photon-limited at moderate spectral
resolution. JWST instruments present the unique ability to work at
more modest spectral resolution and still remain {\it source}-photon
limited. Space-based instruments, overall, will also provide the
most-stable and best-characterized PSFs -- a premium for high
angular-resolution spectrophotometry.

\begin{figure}
\centering
\vspace{5.5cm}
\includegraphics{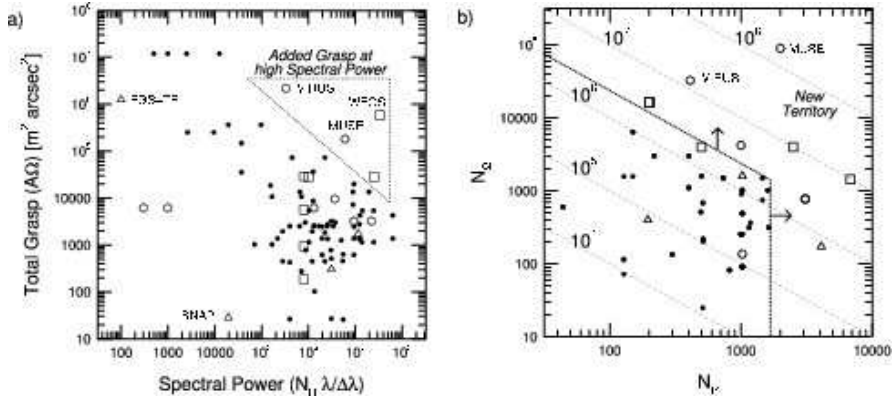}
\caption{Total grasp versus spectral power (a) and the number of
  spatial ($N_\Omega$) versus spectral ($N_R$) resolution element (b)
  including future instruments in Tables 5 and 6: symbols as in
  previous figure.}
\end{figure}

Figure 1.17 reveals where new instruments open up new territory -- both
in added grasp at high spectral power, and simply in more resolution
elements ($N_R \ \times \ N_\Omega$). Of particular note is the thrust toward
instruments with many thousands of spectral resolution elements.
These gains are seen for both ground- and spaced-based instruments, on
both 10m- and 30m-class telescopes.

These gains are made with conceptually conventional grating-dispersed
systems or F-P monochromators.  Clearly there is opportunity for
less-conventional field-widened instruments (such as the
interferometric concepts discussed above), which can amplify both
grasp and spectral power or spectral resolution. Given the relative
novelity of these approaches, they present higher risk, but potentially
higher return, and are best suited for ground-based development.

\medskip
{\it Acknowledgements.}  We would like to thank the IAC and Winter
School organizers, the University of Toronto Department of Astronomy
\& Astrophysics for their gracious hospitality during a sabbatical
year where this work was done, and the NSF for their financial support
of this research (AST-0307417 and AST-0607516).

\end{document}